\documentclass[10pt,journal,compsoc]{IEEEtran}
\usepackage{ifpdf}

\usepackage{xcolor}

\ifCLASSOPTIONcompsoc
\usepackage[nocompress]{cite}
\else
\usepackage{cite}
\fi

\ifCLASSINFOpdf
\usepackage[pdftex]{graphicx}
\else
\usepackage[dvips]{graphicx}
\fi
\usepackage{amsmath}
\usepackage{amssymb}
\usepackage{algorithm}
\usepackage{algorithmic}

\usepackage{array}
\usepackage{tabularx}
\usepackage{multirow}
\usepackage{booktabs}

\usepackage{stfloats}
	\usepackage[hidelinks]{hyperref}
	\hypersetup{
		colorlinks=true,
		linkcolor=blue,
		filecolor=blue,      
		urlcolor=blue,
		citecolor=blue,
	}
	\usepackage{url}
	\usepackage{stmaryrd}
	\usepackage{microtype}

	\usepackage{graphicx}       
	\usepackage{booktabs}       
	\usepackage{multirow}       
	\usepackage{pifont}         
	\usepackage[table]{xcolor} 
	\usepackage{booktabs}
	\usepackage{array}
	
\begin{document}
		%
		\title{SP2UBI: Secure and Privacy-Preserving Usage-Based Insurance}
		%
		%
		%
		%
		
		\author{Mobin~Aghamirkarimi,
			Matin~Aghamirkarimi,
			Farid~Zaredar,
			and~Morteza~Amini}

	\IEEEtitleabstractindextext{%
		\begin{abstract}
			The transition from traditional auto insurance, whose basis is defined by a set of static parameters, like the age of driver and number of accidents, to Usage-Based Insurance (UBI), whose pricing is based on driving behavior, was boosted by Intelligent Transportation Systems (ITS). At the same time, the analysis of high-resolution telematics data might reveal users' behavior and habits, which is associated with significant privacy concerns. The majority of privacy-preserving UBI systems are subject to policyholders' data leaks at some point during the life cycle of their protocol. Moreover, it is challenging to ensure data integrity against possible intentional or unintentional sensor anomalies (like sensor spoofing or malfunctioning) since the approaches used in this case require privacy-compromising audits and do not consider the oracle problem. In order to address the issues, we propose \textit{SP2UBI}, a privacy-preserving UBI solution which guarantees mutual confidentiality. \textit{SP2UBI} collects the telematics data in coarse-grained, statistical form without any spatiotemporal identifiers so that it is impossible to reconstruct fine-grained mobility traces. By utilizing the Torus Fully Homomorphic Encryption (TFHE) scheme, which incurs low computational overhead, computations are performed directly over encrypted data. This way, the insurer calculates risk factors without accessing any sensitive information, while parameters of its risk model are confidential. In order to protect data integrity against sensor level manipulation, \textit{SP2UBI} incorporates a speed verification system assisted by Integrated Sensing and Communications (ISAC) technology which is capable of detecting fraud while preserving user privacy. Experimental evaluation shows that one round of protocol execution takes $41.2\,\text{ms}$, showing that our framework is lightweight. Compared to the other state-of-the-art solutions, \textit{SP2UBI} preserves more privacy guarantees and provides needed functionality for both insurers and policyholders.
			
		\end{abstract}
		
		}

	\maketitle

	\IEEEdisplaynontitleabstractindextext

	%
	\IEEEpeerreviewmaketitle

	\IEEEraisesectionheading{\section{Introduction}\label{sec:introduction}}

	%
	%
	%
	%
	\IEEEPARstart{T}{raditional} vehicle insurance models have historically relied on static demographic data, such as driver age, vehicle model, and accident history. A major limitation of these models is their failure to incentivize safer driving, as they fundamentally ignore real-time, individual driving behavior \cite{Qi2021ScalableDP}. 
	The incorporation of Internet of Things (IoT), Intelligent Transportation Systems (ITS), and Vehicular Ad-hoc Network (VANET) technologies into contemporary transportation resulted in the exchange of real-time information and accurate tracking of routes \cite{Dutta2024ACR}. These advancements made it possible to introduce Usage-Based Insurance (UBI). There are three core paradigms within UBI, namely Pay-As-You-Drive (PAYD), Pay-How-You-Drive (PHYD), and Manage-How-You-Drive (MHYD), that are used to calculate premiums based on dynamic risk and driving behavior \cite{Arumugam2019ASO}.
	
	There are multiple advantages provided by telematics insurance both to users and insurers, and also to society at large. At the societal level, it contributes to improved public safety, reduced traffic congestion, and environmental sustainability by encouraging eco-driving habits and lowering emissions \cite{Soleymanian2017SensorD, Peng2015UsagebasedIS}. For insurers and users, the benefits include accurate risk evaluation, reduction of claim costs, and access to value-added capabilities such as real-time emergency response \cite{Singh2019ABA}. Furthermore, telematics insurance financially incentivizes safer driving, allowing low-risk drivers to reduce their vehicle insurance premiums by up to 30\% \cite{Hndel2014SmartphoneBasedMS}.
	Across regions such as the USA, Europe, and Japan, prominent providers of usage-based vehicle insurance include major companies like Allianz, AXA, Zurich, and Tesla. According to forecasts, by 2033 the global market size will reach \$267.4 billion and will keep a stable Compound Annual Growth Rate (CAGR) of 26.2\% \cite{Allied2023}.
	
	However, for an accurate calculation of risks, there is a need for the continuous tracking of kinematic parameters, including mileage, speed, hard braking, and nighttime driving. Continuous tracking raises significant privacy concerns, acting as a major obstacle to the mass adoption of UBI \cite{BenShahar2023, Quintero2022UsersPC}. Even when collecting only kinematic parameters, sensitive information such as the addresses that drivers frequently visit and commuting routines can be inferred \cite{Wan2018PRIDEAP}.
	
	In order to alleviate these issues, some insurers limit the data gathering process only to non-spatial attributes. However, several empirical studies reveal that the extraction of location trajectories from non-spatial kinematic data is highly feasible. For example, by using only a speed pattern of the car and the origin of the trip, it is possible to predict the destination of 26\% of trips with less than 500 meters precision \cite{Gao2013ElasticPY}. Furthermore, combining speed information with real-time traffic models allows attackers to narrow down the real path to a small set of highly probable candidate routes with a 70\% success rate in the absence of Global Positioning System (GPS) \cite{Zhou2017SpeedBasedLT}. Moreover, time series analysis of brake signals through deep learning enables the successful reconstruction of about 92\% of drivers' main routes \cite{Sarker2023BrakeSignalBasedDL}. Thus, the recent research reveal that there are significant privacy risks even in the case of non-spatial data.
	
	In order to resolve this inherent vulnerability, this paper presents \textit{SP2UBI}, which is a privacy-preserving framework designed specifically for vehicular networks of today. By applying Fully Homomorphic Encryption (FHE)  \cite{gentry2009fully}, in particular, Torus Fully Homomorphic Encryption (TFHE) \cite{chillotti2020tfhe}, the proposed \textit{SP2UBI} performs risk factors calculation and speed verification over the encrypted driving data. Therefore, the suggested protocol meets the needs of risk assessment on the side of the insurer while preserving mutual confidentiality as the insurer calculates the risk of the policyholder without obtaining any telematics, and the policyholder's risk evaluation parameters stay secret to the insurer.
	The main contributions of this research are summarized as follows:
	\begin{itemize}
		\item \textbf{Spatiotemporal Anti-Inference \& Privacy Preservation:} The proposed architecture guards against trajectory tracking and temporal profiling by dissociating the driver's identity from their physical location. At the network edge, anonymous communications prevent intermediate entities from identifying policyholders. On the backend, metadata removal and batch shuffling allow the insurer to evaluate actuarial risks without being able to reconstruct travel routes or daily habits. To ensure mutual privacy, all risk evaluations are carried out entirely over encrypted data using FHE.
		
		\item \textbf{Computation Integrity:} To secure outsourced cloud evaluations against malicious manipulations, we propose a nonce-based Homomorphic Message Authentication Code (HomMAC). This approach guarantees the algebraic integrity of the homomorphic computations while preserving the temporal indistinguishability of the daily records.
		
		\item \textbf{Speed Verification:} To tackle hardware-level sensor spoofing without compromising driver anonymity, \textit{SP2UBI} employs an Integrated Sensing and Communications (ISAC)-assisted validation mechanism. By linking physical ground-truth measurements (e.g., Doppler shifts) to the cryptographic domain, our approach addresses the oracle problem specifically at the speed sensor level. This problem occurs when a compromised Vehicle Speed Sensor (VSS) feeds falsified kinematics into an otherwise secure system. While traditional cryptographic tools blindly sign and authenticate whatever raw data they receive, our framework reliably detects these source-level anomalies in a strictly privacy-preserving manner.
		
		\item \textbf{Practical System Efficiency:} Performance evaluations show that despite the integration of advanced cryptographic primitives, the proposed architecture maintains efficient computational and communication overheads, ensuring its practical viability for real-world deployments.
	\end{itemize}
	
	The remainder of this paper is organized as follows. Section II outlines the cryptographic preliminaries. Section III reviews existing UBI architectures. Section IV introduces the system model. Section V details the proposed \textit{SP2UBI} framework. Section VI presents the privacy and security analysis. Section VII evaluates the system performance. Finally, Section VIII concludes the paper.

	\section{PRELIMINARIES}
	To establish a shared technical foundation for the proposed architecture, this section reviews the primary domain concepts and cryptographic primitives. Specifically, we first categorize existing vehicle telematic insurance paradigms and their data requirements, followed by the mathematical principles of Fully Homomorphic Encryption (FHE), Privacy-Preserving Machine Learning (PPML), and Group Signatures (GS).
	
	\subsection{UBI Models}
	The primary Usage-Based Insurance (UBI) models can be systematically categorized based on their basic risk assessment principle, infrastructure complexity, and the complexity of their analytical approaches. As shown in Table~\ref{tab:ubi_models_comparison}, the area of UBI encompasses three main types of models: Pay-As-You-Drive (PAYD), Pay-How-You-Drive (PHYD), and Manage-How-You-Drive (MHYD). 
	However, despite their huge capabilities in risk profiling, the operational utility of these UBI models depends heavily on the granularity and type of collected telematics data. While basic PAYD schemes rely solely on aggregate odometer readings or periodic distance logs over extended billing cycles, PHYD architectures require multi-dimensional kinematic telemetry. To accurately profile driver behavior, PHYD systems must continuously capture time-series metrics such as vehicle speed, rapid acceleration, hard braking frequencies, and sharp cornering. Furthermore, MHYD frameworks expand these data requirements by fusing kinematic measurements with real-time contextual streams—including environmental conditions, traffic density, and road topography—to enable proactive risk mitigation and immediate driver feedback. Consequently, transitioning from basic PAYD to advanced PHYD and MHYD paradigms shifts the analytical engine from simple periodic distance aggregation to complex, continuous behavioral data processing.
	
	\begin{table*}[htbp]
		\centering
		\caption{Technical and economic comparison of Usage-Based Insurance (UBI) models.}
		\label{tab:ubi_models_comparison}
		\small
		\renewcommand{\arraystretch}{1.2}
		\begin{tabularx}{\textwidth}{@{} >{\raggedright\arraybackslash}p{2.8cm} *{3}{>{\raggedright\arraybackslash}X} @{}}
			\toprule
			\textbf{Dimension} & \textbf{PAYD (Pay-As-You-Drive)} & \textbf{PHYD (Pay-How-You-Drive)} & \textbf{MHYD (Manage-How-You-Drive)} \\
			\midrule
			\textbf{Risk Criteria} & Mileage/Distance & Speed, acceleration, braking, cornering & PHYD metrics $+$ real-time context \\
			\addlinespace
			\textbf{Complexity \& Cost} & Low & Moderate & High \\
			\addlinespace
			\textbf{Analytics} & Simple distance aggregation & Machine learning \& behavioral analysis & Big data \& real-time stream processing \\
			\addlinespace
			\textbf{Key Advantage} & Rewards low-mileage drivers & Incentivizes safe driving behavior & Proactive risk mitigation \& instant intervention \\
			\addlinespace
			\textbf{Main Limitation} & Ignores actual driving behavior & Lacks real-time feedback; sensor-dependent & High computational and infrastructure overhead \\
			\bottomrule
		\end{tabularx}
	\end{table*}
	
	\subsection{Homomorphic Encryption}
	Homomorphic Encryption (HE) represents a major cryptographic scheme that helps solve privacy limitations and allows performing mathematical computations on ciphertexts without prior decryption. Earlier examples of this scheme include the Partially Homomorphic Encryption (PHE) schemes (e.g., RSA \cite{rivest1978} and Paillier \cite{paillier1999}) that can perform only one basic algebraic operation. Formally, $\mathsf{Enc}$ is an encryption function, $\star$ and $\diamond$ stand for the corresponding operations performed on the plaintext space and ciphertext space, respectively. Thus, for a PHE scheme, the homomorphic property holds true with respect to only one operation (either addition or multiplication):
	\[
	\mathsf{Enc}(m_1) \diamond \mathsf{Enc}(m_2) = \mathsf{Enc}(m_1 \star m_2)
	\]
	
	In order to bypass the limitation of a PHE scheme, the first viable Fully Homomorphic Encryption (FHE) scheme \cite{gentry2009fully} was developed and consists of four Probabilistic Polynomial Time (PPT) algorithms $(\mathsf{KeyGen}, \mathsf{Enc}, \mathsf{Dec}, \mathsf{Eval})$. An encryption scheme $\mathcal{E}$ is formally defined as \textit{Fully Homomorphic} if and only if it satisfies two core properties for all constructible circuits $C$:
	\begin{itemize}
		\item \textbf{Correctness:} The evaluation of the scheme over a vector of ciphertexts $\mathbf{c} = (c_1, \dots, c_t)$ representing a vector of plaintexts $\mathbf{m} = (m_1, \dots, m_t)$ results in a correct encryption of the circuit’s output:
		\begin{equation}
			\begin{aligned}
				c^* &\leftarrow \mathsf{Eval}_{\mathcal{E}}(evk, C, \mathbf{c}) \\[1ex]
				&\Rightarrow \mathsf{Dec}_{\mathcal{E}}(sk, c^*) = C(m_1, \dots, m_t)
			\end{aligned}
		\end{equation}
		\item \textbf{Compactness:} The complexity of the decryption algorithm and the size of the evaluated ciphertext $c^*$ do not depend on the depth of the circuit and are bounded by a polynomial in the security parameter $\mathrm{poly}(\lambda)$.  
	\end{itemize}
	
	Homomorphic evaluation of the circuit implies the presence of noise in ciphertexts. The evaluation depth of the Somewhat Homomorphic Encryption (SHE) scheme is limited since noise accumulates in ciphertexts and renders them unusable, while for FHE schemes, this problem is solved by using bootstrapping as proposed by Gentry. Some of the notable lattice-based constructions are the CKKS scheme \cite{cheon2017homomorphic} that performs approximate computing over real and complex numbers, and the TFHE scheme \cite{chillotti2020tfhe}. By using the technique of Programmable Bootstrapping (PBS), TFHE allows controlling the amount of noise in ciphertexts and implementing any non-linear functions such as comparisons, thresholding, and boolean gates. Therefore, TFHE is applicable for exact, privacy-preserving discrete decision-making models.
	
	\subsection{Privacy-Preserving Machine Learning}
	Privacy-Preserving Machine Learning (PPML) \cite{Taubert2026APC} involves numerous cryptographic and distributed approaches such as Federated Learning, Multi-Party Computation (MPC), and HE aimed at processing the information without revealing the confidential original data. In the specific context of the FHE approach, PPML helps to convert analytical models and multi-parameter risk-assessment algorithms into efficient computational circuits ready for secure evaluation. For the realization of this conversion in the TFHE scheme \cite{chillotti2020tfhe}, precise parameter configuration becomes necessary since it allows controlling noise growth and the multiplicative depth of the circuit to decrease the necessity of using resource-intensive bootstrapping procedures. Additionally, a unique property of TFHE is the ability to perform PBS \cite{Chillotti2021ProgrammableBE}, allowing the homomorphic evaluation of non-linear functions (like thresholding and step functions), which are required for more sophisticated data analysis.
	
	Considering the inherent limitations of FHE computational stacks regarding the processing of limited bit-widths, the quantization of parameters and intermediate variables becomes a necessary condition. The quantization is typically done with the help of Quantization-Aware Training (QAT), which incorporates discrete restrictions into the gradient optimization process, or Post-Training Quantization (PTQ) based on post-hoc calibration. Eventually, the optimized model will be mapped to an equivalent homomorphic circuit, inheriting the IND-CPA security guarantees of the underlying encryption scheme.
	
	\subsection{Group Signatures}
	Whereas FHE and PPML techniques efficiently ensure telematics data privacy when processed, proving the authenticity and integrity of the data origin before applying the cryptographic transformations poses an essential problem. Group Signature (GS) \cite{chaum-group-sig} is a modern cryptographic primitive solving this problem since it enables group members to sign messages anonymously on behalf of the entire group. Unlike traditional digital signature schemes where the identity of the signer is revealed to everybody, the GS scheme allows verifying the fact that the message was indeed signed by one of the authorized parties without disclosing the signer's specific identity \cite{Zhang2024MessageLG}. Thus, the vehicle identity becomes hidden from the adversaries.
	
	In order to balance high levels of user privacy with system accountability, the GS protocol incorporates a trusted entity known as the Group Manager (GM) \cite{bellare-gs-foundations}. In cases of dispute, fraud, or malicious behavior, the GM utilizes a unique tracking key capable of revoking the anonymity and revealing the true identity of a rogue signer. Such conditional privacy, combined with definitive traceability, makes group signatures exceptionally valuable for securing data integrity in resource-constrained and dynamic IoT systems and smart vehicular architectures \cite{boneh-bbs-short-group-sig}.

	\section{RELATED WORK}
	To contextualize the proposed framework within the broader landscape of telematics insurance, existing privacy-preserving UBI solutions can be grouped based on their underlying architectural design and the methodologies employed to balance data integrity with user anonymity. Specifically, the related literature is examined across three methodological categories: (i) hardware-based security and tamper resistance, (ii) distributed and blockchain-based approaches, and (iii) noise-addition and data perturbation techniques. In the following, the operational concepts, trust assumptions, and practical limitations associated with each category are discussed.
	
	\subsection{Hardware-Based Security and Tamper Resistance}
	To minimize the reliance on potentially malicious vehicular sensors and to protect telematics data from being deliberately altered, various hardware solutions have been actively considered. Due to the inability of insurers to implicitly trust the native Electronic Control Units (ECUs) in a vehicle, these solutions generally require installing independent hardware modules, such as telematics black boxes containing a separate sensor suite. Despite early claims about the tamper-proof nature of these systems, in practice, they can only be implemented as either tamper-resistant or, at best, tamper-evident. These architectures perform some form of localized processing to prevent insurers from accessing the raw telematics data. For instance, Troncoso et al. \cite{Troncoso2007PriPAYDPP} introduced \textit{PriPAYD}—a privacy-oriented Pay-As-You-Drive architecture where insurance premiums are calculated locally in the black box based on information obtained from GPS, Global System for Mobile Communications (GSM), and digital maps. Though \textit{PriPAYD} provides strict data minimization since the only transmitted piece of data is an aggregated premium, its adoption is hindered by the high expenses of making the hardware tamper-resistant \cite{Wan2018PRIDEAP}, significant computational overhead for localized map-matching, and the problem of ensuring hardware integrity in light of possible physical vulnerabilities. 
	In order to address the computational overhead associated with localized map-matching, Popa et al. \cite{Popa2009VPrivPP} introduced the \textit{VPriv} system, which integrates cryptographic commitments, secure multi-party computation, and interactive zero-knowledge proofs (ZKPs) into the design of a tamper-resistant transponder. The premium calculation in \textit{VPriv} is performed on the server side over anonymized spatiotemporal logs, thus guaranteeing raw data confidentiality. However, the intensive use of these interactive cryptographic primitives for continuous, large-scale telematics streams severely limits their scalability due to significant computational latency and communication overhead. Also, to ensure compliance and detect evasion, \textit{VPriv} relies on random physical roadside inspections of license plates and location points \cite{Wan2018PRIDEAP}, which inevitably introduces secondary privacy risks. 
	At the same time, even securing the hardware housing does not preclude the presence of threats at the sensor layer. As demonstrated by Fu et al. \cite{Fu2018RisksOT}, sensors can still be spoofed and attacked via transduction (signal injection) even without invasive physical access, since adversaries can exploit the fundamental physical properties of these components to manipulate analog outputs. Ultimately, while hardware-based solutions increase the integrity of telematics through localized processing, their practical viability is hampered by manufacturer-dependent trust assumptions, the privacy degradation inherent in physical spot-checking, and the high expenses associated with widespread deployment and maintenance.
	
	\subsection{Distributed and Blockchain-Based Approaches}
	
	Several systems have utilized decentralized architectures to protect vehicular telematics data. For instance, \textit{PRIDE}, a privacy-driven framework that integrates blockchain technology and homomorphic encryption in a vehicle-cloud-blockchain setting, was proposed by Wan et al.~\cite{Wan2018PRIDEAP}. This solution encrypts multi-dimensional driving vectors into invertible matrices, then anchors the Merkle tree roots generated from them onto the blockchain network, allowing smart contracts to compute acceleration-to-speed risk factors using encrypted data. Despite the fact that \textit{PRIDE} ensures computations over ciphertexts, its reliance on multi-dimensional driving vectors encoded into binaries requires extensive storage capacity~\cite{Sahu2024BlockchainAM}. In addition, \textit{PRIDE} faces significant privacy issues due to continuous data stream transmissions that might expose the underlying spatiotemporal behavioral profiles of the driver~\cite{Qi2021ScalableDP}.
	
	In order to address the issue of tracking, Singh et al.~\cite{Singh2019ABA} proposed a multi-tier architecture that involves the integration of both public and private blockchains with a Proof-of-Work (PoW) consensus protocol. In this framework, Traffic Authority-controlled Roadside Units (RSUs) operate the public blockchain network while dynamically changing short-lived pseudonyms are assigned to vehicles through the private issuer blockchain network for performing periodic transactions. Even though the address rotation approach reduces tracking vectors and improves anonymity, relying on RSUs as mining nodes for an energy-consuming PoW consensus incurs prohibitive computational overheads, limiting its viability for large-scale vehicular networks.
	
	In regard to the problem of raw data exposure, Qi et al.~\cite{Qi2021ScalableDP} introduced the \textit{DUBI} system that exploits Pedersen commitments, non-interactive zero-knowledge proofs (NIZKPs), and Merkle trees. In \textit{DUBI}, the driving metrics are kept in encrypted form as commitments in a cloud broker while their Merkle tree roots are anchored to the blockchain network to perform insurance premium calculations through smart contracts. Unfortunately, this framework still requires partial data disclosure during random spot checks or post-collision inspections, thereby introducing localized privacy leakage vectors. Similar to \textit{DUBI}, Huang et al.~\cite{Huang2023BlockchainAssistedPC} presented a consortium blockchain solution that integrates the Paillier cryptosystem, Fujisaki-Okamoto commitments, and ZKPs. In this system, fraud detection is modeled as a recursive inspection game to achieve a Nash equilibrium. Though this system is designed to strike a balance between transparency and fraud prevention through an independent third-party auditor, its reliance on a fully trusted auditing entity introduces privacy vulnerabilities during the verification phase, as the auditor gains visibility into the raw driving data.
	
	To provide data integrity without centralized raw telematics collection while maintaining auditability, Yi et al.~\cite{Yi2022CCUBIAC} proposed \textit{CCUBI}, a cross-chain insurance framework. In this work, instead of transmitting fine-grained variables, vehicles commit cryptographic proofs of multi-dimensional driving indicators to the blockchain, leveraging random blinding factors to prevent the brute-force resolution of discrete logarithms. Meanwhile, RSUs conduct random spot checks against physical traffic observations. Even though this solution ensures data trustworthiness, its scalability is restricted by the finite storage capacity of RSUs for short-term data retention, alongside privacy issues related to localized data exposure during roadside verification. As for edge-based intelligence, Sahu et al.~\cite{Sahu2024BlockchainAM} proposed a framework called \textit{BE-VIP} that combines blockchain, local analytical models, and computer vision. Local processing of the driving data occurs at the vehicle level via the use of logistic regression to calculate a safety score. To reduce the cost of on-chain storage, only this final index, alongside the cryptographic hashes of Fujisaki-Okamoto commitments and accident images stored on the InterPlanetary File System (IPFS), are anchored to the ledger. Although such a framework provides automated claims processing and preserves the confidentiality of raw telematics at the vehicular tier, the reliability of the safety score calculation strictly depends on the physical sensor layer. Hence, this framework is susceptible to data integrity compromises if the speed or kinematic sensors are manipulated or spoofed prior to the local processing of the data. Consequently, despite enhancing data integrity and transparency, these distributed frameworks remain constrained by storage and scalability bottlenecks, auditing-related privacy trade-offs, and a critical vulnerability to physical-layer sensor anomalies.
	
	\subsection{Noise-Addition and Data Perturbation Techniques}
	The approach by Zhou et al.~\cite{Zhou2019LocationPI}, known as \textit{Pri-UBI}, represents the integration of cryptography with the synchronous injection of random noise in order to simultaneously ensure location privacy and prevent data fabrication. The design matches an in-vehicle On-Board Diagnostics (OBD-II) telematics device with smartphone sensors by injecting identical random variables into both data streams before their submission. Because these random variables are specifically designed to sum to zero over a designated epoch, this mechanism ensures that the fine-grained profiling of an individual is concealed, while at the same time allowing the calculation of precise insurance risk scores over ciphertexts. For the purpose of preventing fraud, the authors use a probabilistic Usage Data Audition scheme that can be considered an inspection game. By probabilistically cross-checking the encrypted data obtained from the two sources, calculating the difference between the streams eliminates the shared noise and isolates any unauthorized data alterations. However, despite being able to balance privacy and fraud prevention through dual-source correlation, the architecture remains limited by the need for constant local synchronization between the smartphone and the OBD-II system for real-time key exchange. Furthermore, the selective application of zero-sum noise exclusively to continuous metrics leaves unperturbed discrete telematics features, such as timestamped hard braking events, susceptible to trajectory reconstruction attacks.

	\section{Problem Definition and Overall Proposed Solution}
	To establish a clear technical context, this section first articulates the fundamental privacy and physical-layer security challenges in modern telematics insurance. Following this problem formulation, the physical system architecture of the proposed \textit{SP2UBI} framework is presented and the trust assumptions governing the involved entities are outlined.
	
	\subsection{Problem Definition}
	While advanced UBI paradigms, specifically PHYD and MHYD, offer significant benefits in dynamic risk profiling, their real-world deployment is impeded by two unresolved structural vulnerabilities:
	
	\begin{itemize}
		\item \textbf{Spatiotemporal Privacy Leakage and Outsourced Computation Integrity:} Sending continuous, fine-grained driving metrics—such as vehicle speed, rapid acceleration, and braking frequency—to external cloud servers creates severe privacy risks. Even without explicit GPS coordinates, an untrusted server can analyze time-series kinematic patterns to reconstruct a driver's daily commute, uncover private addresses, and build a detailed behavioral profile. This presents a dual challenge. First, the system must allow the insurer to evaluate actuarial risk models over encrypted, unordered data without ever exposing the underlying telemetry. Second, because these calculations are offloaded to third-party cloud providers, the insurance provider needs a dependable mechanism to verify that its outsourced model is executed correctly and honestly, guarding against any computational manipulation.
		
		\item \textbf{The Physical-Layer Oracle Problem in Speed Telemetry:} To prevent data tampering, most existing vehicular protocols rely on software-layer cryptography, such as digital signatures and zero-knowledge proofs generated inside an OBU. The drawback is that these tools simply assume the measurements coming from internal sensors are correct to begin with. This exposes a classic oracle problem: if the VSS suffers from an unintentional malfunction or is physically tampered with before the signal is digitized, the OBU will inevitably authenticate erroneous data. Vehicle speed is a highly important parameter for evaluating driver risk factors. However, conventional methods for catching sensor anomalies or fraud usually require inspecting fine-grained, plaintext driving logs or relying on random inspections—practices that severely infringe on user privacy. Therefore, a trustless, physical-layer speed verification mechanism is required that can independently detect sensor errors without relying on internal vehicular telemetry and without compromising driver anonymity.
	\end{itemize}
	
	\subsection{Proposed Solution Architecture}
	The architecture of the proposed \textit{SP2UBI} framework comprises four primary entities: the On-Board Unit (OBU), the Roadside Unit (RSU), the Telematics Service Provider (TSP), and the Insurer. The hierarchical interactions among these entities are illustrated in Fig.~\ref{fig:entities}. The specific operational roles of each entity are defined as follows:
	
	\begin{figure}[htbp]
		\centering
		\includegraphics[width=0.9\columnwidth]{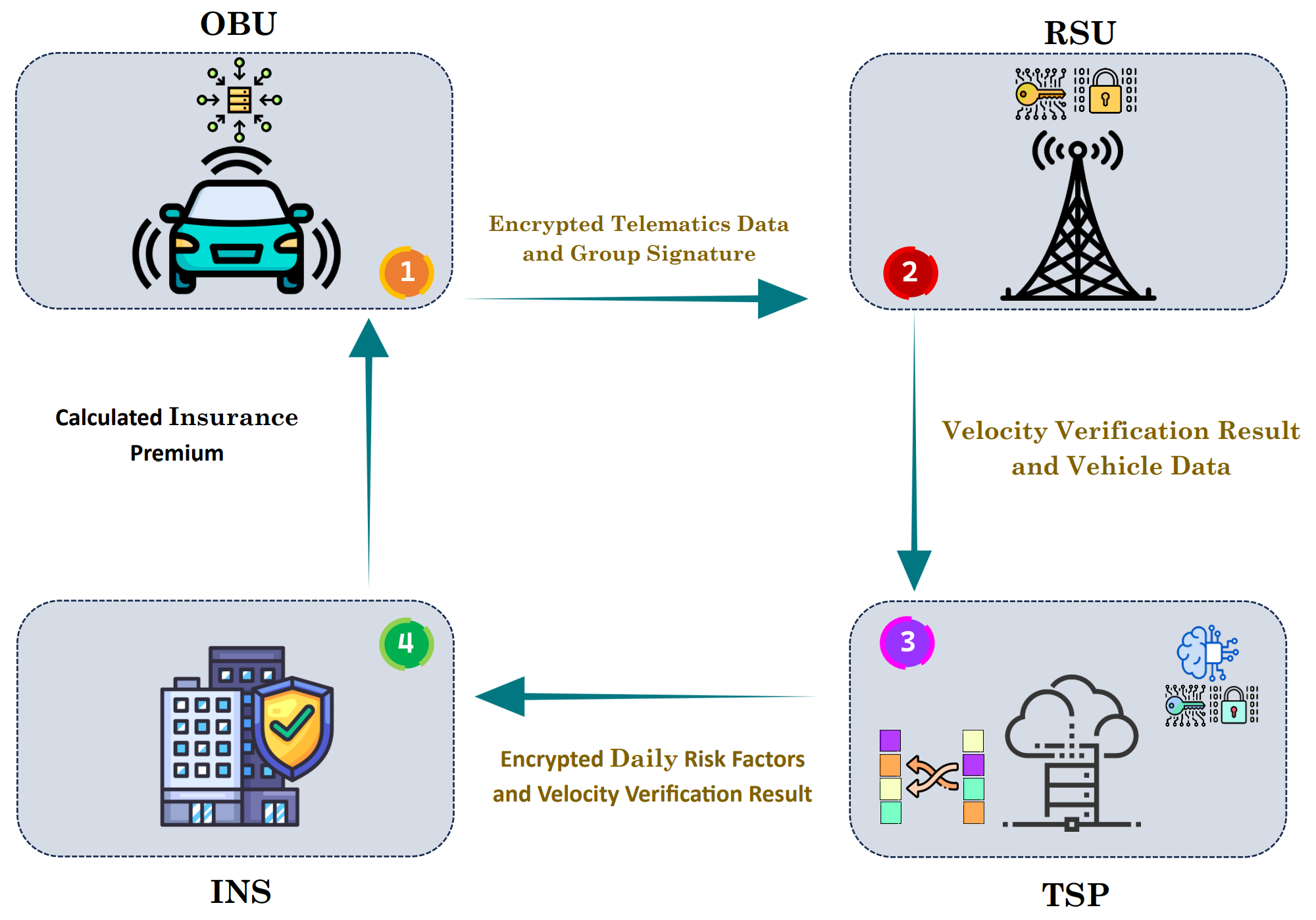}
		\caption{Architecture of the system components and their interactions within the SP2UBI framework.}
		\label{fig:entities}
	\end{figure}
	
	\begin{itemize}
		\item \textbf{On-Board Unit (OBU):} This unit is responsible for interfacing with the physical sensors of the vehicle to collect coarse-grained driving telematics and instantaneous speed. While the cryptographic processor (Trusted Execution Environment (TEE) / Hardware Security Module (HSM)) inside the OBU is trusted, the collected telemetry data from the sensors can still be susceptible to hardware spoofing. In order to protect the confidentiality of the data and create a basis for verifiable computation, the OBU encapsulates driving metrics together with temporal nonces and cryptographic tags (HomMAC), encrypts the payload using FHE under the insurer's public key, and authenticates the transmission with a group signature. The OBU securely keeps a local plaintext copy of the logs up until the end of the billing epoch in case of potential disputes.
		
		\item \textbf{Roadside Unit (RSU):} Serves as the local ISAC gateway in the vehicular ad-hoc network. The main objective of the RSU is the physical verification of the transmitted speed data. After receiving the uplink message, the RSU measures the physical ground-truth speed based on the Doppler shift and estimates the reliability of the channel. Then, it uses a lightweight, mixed-domain homomorphic verification circuit that computes directly over the encrypted sensor measurements and the physical measurement. The RSU forwards the authenticated, encrypted driving payload with the verification discrepancy to the TSP. 
		
		\item \textbf{Telematics Service Provider (TSP):} Serves as the primary computational and obfuscation intermediary. It receives the risk factor assessment model created by the insurer in the form of a linear FHE circuit and homomorphically evaluates the model over the encrypted telematics data and cryptographic tags. In order to preserve the multidimensional anonymity of the data, the TSP actively removes any localized geographic identifiers and RSU identifiers, which allows the prevention of spatial inference. Additionally, it securely aggregates the daily evaluated payloads and applies a cryptographically secure random permutation (shuffling) at the end of the epoch in order to prevent temporal inference and chronological profiling before sending the batch to the insurer.
		
		\item \textbf{Insurer:} Receives the batch of shuffled, anonymized, and encrypted evaluations from the TSP. The objective of the insurer is restricted to decrypting the received aggregate metrics. Prior to computing the insurance premium, the insurer must mathematically verify the cryptographic execution integrity of the TSP (via HomMAC algebraic congruence) and also check whether the decrypted speed discrepancy is within the acceptable physical tolerances. This architecture allows the insurer to maintain risk evaluation utility and auditability while remaining entirely oblivious to the policyholder's raw spatiotemporal trajectories.
	\end{itemize}
	\subsection{Threat Model}
	Under the assumption of the threat model in the proposed scheme, the OBU is considered trusted with regard to executing its designated operations. Although the OBU is trusted, it receives data from the physical sensor layer, which is taken as an untrusted domain subject to anomalies in the environment, hardware degradation, or even external spoofing (the oracle problem).
	
	However, both the RSU edge infrastructure and the TSP are considered as honest-but-curious. Under normal operating conditions, they follow all the protocol specifications but might try to learn private trajectories from the intercepted ciphertexts. It is important to note that a significant advantage of the proposed \textit{SP2UBI} architecture is its resilience beyond standard semi-honest assumptions. In cases where an active adversary breaks into the TSP and tries to manipulate the ciphertexts or forge evaluations, the proposed verification mechanism allows the insurer to reliably detect these anomalies and guarantees execution integrity.
	
	Lastly, the insurance provider is assumed to be honest-but-curious. Even though accurate risk profiling is required to compute fair premiums, its access is cryptographically limited to only the decrypted ``Risk Factor'' and the ``Speed Verification'' status. Access to raw vehicular data is mathematically impossible, ensuring strict policyholder privacy without compromising the utility of the UBI model.
	
	\section{DETAILED DESCRIPTION OF SP2UBI}
	This section describes the details of the proposed \textit{SP2UBI} framework. To provide a clear structural progression, the presentation separates the underlying techniques from the operational protocol workflow. Specifically, the first three subsections outline the core cryptographic and physical-layer building blocks: the privacy-preserving risk factor calculation, the Integrated Sensing and Communications (ISAC)-assisted speed verification mechanism, and the homomorphic authentication scheme designed for temporal privacy. Following the introduction of these foundational tools, the final subsection describes the complete end-to-end protocol, detailing the step-by-step execution phases and data exchanges across all participating entities.
	
	\subsection{Risk Factor Calculation}
	\label{subsec:risk_factor}
	
	To establish the risk assessment baseline in \textit{SP2UBI}, the On-Board Unit (OBU) aggregates raw vehicular metrics into a coarse-grained telematics feature vector. Rather than transmitting privacy-sensitive, fine-grained event logs—such as individual braking timestamps which expose trajectory correlates—the OBU computes cumulative statistical summaries over a designated temporal interval. For a given day $d$, this daily feature vector, denoted generally as $\mathbf{x}^{(d)}$, is formally defined as:
	\begin{equation}
		\mathbf{x}^{(d)} = \big( \mathit{ID}, x_{\text{hb}}^{(d)}, x_{\text{sv}}^{(d)}, x_{\text{spd}}^{(d)}, x_{\text{mi}}^{(d)}, x_{\text{rh}}^{(d)} \big)
		\label{eq:M_insurance}
	\end{equation}
	where $\mathit{ID}$ represents the unique private policyholder's identifier. The remaining vector elements correspond to key behavioral and contextual metrics for that specific day: $x_{\text{hb}}^{(d)}$ denotes the cumulative frequency of hard braking events defined as deceleration $> 7.7~\mathrm{mph/s}$\footnote{Nationwide SmartRide Program. [Online]. Available: \url{https://www.nationwide.com/}}, $x_{\text{sv}}^{(d)}$ represents the frequency of speed violations $> 80~\mathrm{mph}$\footnote{Kia Usage-Based Insurance. [Online]. Available: \url{https://owners.kia.com/content/owners/en/usage-based-insurance.html}}, $x_{\text{spd}}^{(d)}$ is the average trip speed, $x_{\text{mi}}^{(d)}$ is the total accumulated mileage, and $x_{\text{rh}}^{(d)}$ indicates the driving duration during high-risk temporal windows, such as 23:00--05:00\footnote{Allstate Drivewise. [Online]. Available: \url{https://www.allstate.com/}}.
	
	This flexible aggregation interval is highly configurable to align with the standard risk-profiling requirements of commercial insurers. As a practical baseline, a 24-hour epoch is selected to maximize data fault tolerance. This setup ensures that localized telemetry corruption or short-term network invalidation events do not compromise extensive volumes of historical metrics, while effectively protecting the driver's fine-grained spatiotemporal trajectories from passive infrastructure exposure.
	
	In practice, the TFHE framework is utilized to process vehicular metrics entirely within the encrypted domain. This secure computation is controlled by an asymmetric FHE key management system: the public encryption key ($\mathit{pk}_{\text{Ins}}$) is distributed to the OBUs for payload encryption; the evaluation key ($\mathit{evk}_{\text{Ins}}$) is provisioned to the Telematics Service Provider (TSP) to allow homomorphic computation; and the secret decryption key ($\mathit{sk}_{\text{Ins}}$) is held exclusively by the insurer. Specifically, any ciphertext encrypted under a public key $\mathit{pk}$ is formally denoted as $\llbracket \cdot \rrbracket_{\mathit{pk}}$. The risk evaluation model $\mathcal{F}_{\text{Ins}}$ of the insurer's choice is compiled into a linear FHE evaluation circuit and outsourced to the TSP. The TSP performs computations on the encrypted input vector $\llbracket \mathbf{x}^{(d)} \rrbracket_{\mathit{pk}_{\text{Ins}}}$ to obtain the encrypted risk result, which ensures that real vehicular data remains strictly inaccessible to the processing cloud. Crucially, the private identifier $\mathit{ID}$ is securely embedded inside the encrypted payload. As a consequence, the TSP performs the computation blindly and cannot correlate specific ciphertexts with actual users, thereby mitigating potential trajectory tracking vectors.
	
	To compute the policyholder's daily risk factor without compromising data confidentiality, the proposed framework processes these encrypted vehicular metrics utilizing the insurer's specific risk assessment model. Due to the proprietary nature of commercial evaluation architectures, the insurer's risk assessment is abstracted as a generalized function $\mathcal{F}_{\text{Ins}}$ that maps the encrypted behavioral features to an encrypted daily risk factor score, denoted as $\llbracket \mathcal{R}_{\text{Ins}}^{(d)} \rrbracket_{\mathit{pk}_{\text{Ins}}}$:
	\begin{equation}
		\begin{split}
			\llbracket \mathcal{R}_{\text{Ins}}^{(d)} \rrbracket_{\mathit{pk}_{\text{Ins}}} = \mathcal{F}_{\text{Ins}} \Big( 
			&\llbracket x_{\text{hb}}^{(d)} \rrbracket_{\mathit{pk}_{\text{Ins}}}, \llbracket x_{\text{sv}}^{(d)} \rrbracket_{\mathit{pk}_{\text{Ins}}}, \llbracket x_{\text{spd}}^{(d)} \rrbracket_{\mathit{pk}_{\text{Ins}}}, \\
			&\llbracket x_{\text{mi}}^{(d)} \rrbracket_{\mathit{pk}_{\text{Ins}}}, \llbracket x_{\text{rh}}^{(d)} \rrbracket_{\mathit{pk}_{\text{Ins}}} \Big)
		\end{split}
		\label{eq:RiskFactor}
	\end{equation}
	
	Since the data is already encrypted under the TFHE scheme, this framework inherently allows for evaluating quite complicated machine learning models through its PBS. Despite this capability, obtaining the exact structural parameters of commercial insurance models remains infeasible due to high confidentiality standards. For this reason, to set up a replicable baseline and considering the operational assumptions of existing telematics studies \cite{Liu2017ADB}, the risk factor evaluation model is abstracted as a linear scoring function. This realistic assumption is consistent with standard industry practices (e.g., Generalized Linear Models) and fits the algebraic conditions of the designed execution verification scheme (detailed in Section~\ref{subsec:hommac_temporal}). However, the built-in encryption layer guarantees the flexibility for insurers to securely integrate non-linear operations via PBS, accommodating more complex risk evaluation models in subsequent industrial applications.
	
	Finally, once the billing period is over, the TSP randomly shuffles the global pool of encrypted daily risk records before delivering them in a batch to the insurer. As a result, upon decryption, the insurer reveals the $\mathit{ID}$ and accesses the daily factors in plaintext, preventing both spatial tracking by the edge and continuous temporal profiling by the cloud.
	
	\subsection{ISAC-Assisted Speed Verification with Reliability Awareness}
	\label{subsec:velocity_verification}
	Many existing techniques face certain difficulties associated with data provenance. For instance, the \textit{PRIDE} scheme \cite{Wan2018PRIDEAP} verifies data solely based on the mathematical coherence between speed and acceleration. This technique remains vulnerable to hardware-based spoofing attacks, since an attacker can systematically construct artificial telemetry that satisfies these theoretical criteria.
	
	Similar to that, the solution suggested by Huang et al.~\cite{Huang2023BlockchainAssistedPC} incorporates OBU digital signatures and ZKPs yet overlooks the physical reality of the car itself. Usually, cryptographic techniques are applied to secure data only after it has been digitized, thereby exposing the fundamental oracle problem. If the VSS is compromised at the hardware level, the OBU authenticates the fake data and subsequently generates computationally valid ZKPs for the fabricated metrics. What is more, in modern automotive architectures, the odometer physically relies on VSS readings to accumulate miles; thus, modifying the speed results proportionally changes the total mileage. Hence, any physical auditing mechanisms based strictly on comparing distances can be easily bypassed.
	
	Since speed sensor telemetry serves as the fundamental basis for calculating driving risk factors, as mentioned in Section~\ref{subsec:risk_factor}, verifying its physical integrity is paramount. In order to avoid the circular trust and data association vulnerabilities inherent to separated radar and communication solutions, the \textit{SP2UBI} framework makes use of ISAC \cite{liu2022integrated}. In this paradigm, the sensing and communication capabilities share the exact same hardware infrastructure and employ a single Radio Frequency (RF) waveform. Extracting the kinematics of the vehicle directly from the Doppler shift and delay of the communication echoes allows ISAC to natively bind the physical ground-truth of the vehicle with its cryptographic digital identity. As a result, there is no need to rely on untrusted self-reports, such as on-board VSS readings or GPS coordinates, effectively bypassing the oracle problem.
	
	The verification process relies on the deterministic physical-layer sensing of the uplink communication signal. Once the telematics packet is sent to the edge for a given day $d$, the OBU retrieves the instantaneous speed of the vehicle, denoted as $v_{\text{sen}}^{(d)}$, directly from the onboard sensors. In order to preserve privacy, the OBU homomorphically encrypts this reading using the dedicated verification key ($\mathit{pk}_{\text{SV}}$) and transmits the encrypted ciphertext $\llbracket v_{\text{sen}}^{(d)} \rrbracket_{\mathit{pk}_{\text{SV}}}$ embedded inside an ISAC waveform (for example, an Orthogonal Frequency Division Multiplexing (OFDM) signal).
	
	Concurrently, upon receiving this uplink transmission via orthogonal resources (e.g., Orthogonal Frequency Division Multiple Access (OFDMA) / Space Division Multiple Access (SDMA)), the RSU decodes the communication payload while simultaneously extracting physical sensing parameters from the exact same electromagnetic signal. Because the transmitted data packet and the physical sensing target originate from a singular electromagnetic emission, data association is intrinsically guaranteed, overcoming the target-matching ambiguities typical of traditional Frequency Modulated Continuous Wave (FMCW) radars \cite{Patole2017AutomotiveRA}. Based on the ISAC vehicular model \cite{liu2022integrated}, the RSU directly extracts the Doppler shift ($f_D$) and the Angle of Arrival ($\theta$). The vehicle's true physical speed ($v_{\text{isac}}^{(d)}$) is geometrically compensated at the edge:
	\begin{equation}
		v_{\text{isac}}^{(d)} = \frac{c f_D}{f_c \cos(\theta)}
		\label{eq:isac_velocity}
	\end{equation}
	where $c$ is the speed of light and $f_c$ is the carrier frequency. Unlike two-way monostatic radar configurations, the one-way uplink transmission natively excludes the factor of $2$ in the denominator, and standard pilot-based compensation is applied to isolate $f_D$ from asynchronous Carrier Frequency Offsets (CFO). This unified ISAC approach essentially abstracts the physical infrastructure into a reliable, GPS-independent speed oracle, allowing the framework to focus exclusively on cryptographic validation.
	
	In real-world traffic scenarios, the extracted measurement may be compromised by degraded channel conditions (e.g., severe multipath fading or Non-Line-of-Sight reflections) or numerical instability when the vehicle passes directly perpendicular to the RSU ($\theta \to 90^\circ, \cos(\theta) \to 0$). To prevent false-positive penalization of honest drivers, the RSU evaluates the physical-layer reliability before executing any cryptographic operations. Using the measured Signal-to-Noise Ratio (SNR) $\gamma^{(d)}$ and the predefined threshold $\gamma_{\text{th}}$, a Boolean reliability indicator $\Phi_{\text{rel}}^{(d)} \in \{0, 1\}$ is defined, where $\Phi_{\text{rel}}^{(d)} = 1$ if and only if the channel satisfies three physical criteria: sufficient SNR ($\gamma^{(d)} \ge \gamma_{\text{th}}$), Line-of-Sight (LoS) dominance, and a geometrically stable Angle of Arrival ($|\cos(\theta)| \ge \delta_{\theta}$) to avoid singularity. Based on this indicator, the RSU dictates the subsequent operation to produce the daily verification result $\mathcal{R}_{\text{SV}}^{(d)}$:
	
	\begin{equation}
		\mathcal{R}_{\text{SV}}^{(d)} = 
		\begin{cases} 
			\llbracket v_{\text{sen}}^{(d)} \rrbracket_{\mathit{pk}_{\text{SV}}} \ominus v_{\text{isac}}^{(d)} & \text{if } \Phi_{\text{rel}}^{(d)} = 1 \\[1.5ex]
			\text{FLAG}_{\text{uncert}} & \text{if } \Phi_{\text{rel}}^{(d)} = 0 
		\end{cases}
		\label{eq:result_sv}
	\end{equation}
	
	If the measurement is considered reliable ($\Phi_{\text{rel}}^{(d)} = 1$), in order to reduce computational complexity at the edge infrastructure, the RSU applies a light-weight verification circuit $\mathcal{F}_{\text{SV}}$ homomorphically using the evaluation key $\mathit{evk}_{\text{SV}}$, processing values in the mixed domain of ciphertext and plaintext. Given the discrete nature of FHE, both the ciphertext and plaintext values are represented as integers by quantizing the encrypted sensor data and ISAC measurement using the same universally shared scaling factor. Instead of performing the expensive homomorphic comparison of magnitudes, the RSU calculates the difference of the values using the homomorphic subtraction operation ($\ominus$). However, if the channel or geometry is unreliable ($\Phi_{\text{rel}}^{(d)} = 0$), the RSU completely skips the use of the homomorphic circuit, producing the deterministic $\text{FLAG}_{\text{uncert}}$ value indicating a lack of physical-layer confidence.
	
	To integrate seamlessly with the underlying privacy-preserving protocol architecture, the RSU sends the result of the verification process $\mathcal{R}_{\text{SV}}^{(d)}$ along with the encrypted telematics message to the TSP. As defined in the system model, the TSP systematically removes the geolocation metadata and shuffles the batch of daily telematics messages before sending it to the insurance provider at the end of the billing epoch. Upon decryption of the batch entries, if the $\text{FLAG}_{\text{uncert}}$ value is encountered by the insurance provider, then the specific daily verification is ignored neutrally without any penalty to the driver. Otherwise, the verification result is decrypted by the insurance provider using the secret verification key ($\mathit{sk}_{\text{SV}}$), and the absolute error is compared against the defined tolerance threshold ($\epsilon$):
	\begin{equation}
		\big| \mathit{Dec}\bigl(\mathit{sk}_{\text{SV}}, \, \mathcal{R}_{\text{SV}}^{(d)}\bigr) \big| \le \epsilon
		\label{eq:threshold_check}
	\end{equation}
	In which $\epsilon$ denotes the policy-defined margin, taking into account the tolerances of the physical hardware and inherent estimation error of the ISAC technique, bounded from below by the Cramér-Rao Lower Bound (CRLB) \cite{liu2022integrated}. If the threshold is exceeded, then the temporal epoch is marked as potentially affected by hardware spoofing. This fault-tolerant architecture implicitly exploits the ISAC to ensure trustless data validation, while maintaining the driver privacy and smartly offloading absolute decision-making to the trusted insurance provider.
	
	\subsection{Temporal-Privacy Preserving Homomorphic MAC for Daily Evaluations}
	\label{subsec:hommac_temporal}
	
	As discussed in Section~\ref{subsec:risk_factor}, the insurer's evaluation model is assumed to be a linear function. To achieve execution integrity without revealing the timing of daily evaluations inside the billing epoch $t$, a nonce-based Homomorphic Message Authentication Code (HomMAC) is proposed. Let $D$ be the number of days in epoch $t$. During the offline phase, the OBU and the insurer generate and agree on a shared symmetric secret key $K$.
	
	For each day $d \in \{1, \dots, D\}$, once the daily coarse-grained driving data vector $\mathbf{x}^{(d)} = \big( \mathit{ID}, x_{\text{hb}}^{(d)}, x_{\text{sv}}^{(d)}, x_{\text{spd}}^{(d)}, x_{\text{mi}}^{(d)}, x_{\text{rh}}^{(d)} \big)$, as defined in Equation~(\ref{eq:M_insurance}), is ready for transmission, the OBU initiates the cryptographic binding process using a distinct set of parameters. The structural parameters ($\alpha_t$, $r_i$, and $p$) are deterministically derived from the shared key $K$ or pre-agreed. Meanwhile, the daily nonce ($u_d$) is generated independently on the fly. The operational roles of these parameters are structured as follows.
	
	Using a cryptographic Pseudorandom Function (PRF) denoted as $F_K$ keyed by the shared secret $K$, the OBU generates an epoch-specific MAC base scalar $\alpha_t \leftarrow F_K(t \parallel 0)$ to bind the evaluation to the current billing cycle $t$. It then applies the same PRF to compute input-specific base blinding factors $r_i \leftarrow F_K(t \parallel i)$, which act as spatial masks for the individual driving features. While the insurer is authorized to see only the final evaluated scalar risk score ($\mathcal{R}_{\text{Ins}}^{(d)}$), omitting $r_i$ would leave the algebraic structure of the tags unmasked. In such a scenario, an honest-but-curious insurer could exploit its knowledge of the linear model weights, set up a system of equations across the shuffled batch, and solve for the underlying raw driving metrics ($x_{\text{hb}}^{(d)}, x_{\text{sv}}^{(d)}, \dots$). Extracting these fine-grained features violates core data minimization principles. It also exposes the policyholder's exact daily behavioral fingerprint, which directly defeats the privacy objectives of the framework.
	
	To constrain the MAC space and prevent integer overflow during homomorphic evaluations, all computations are securely bounded by a large prime modulus $p$. Within this finite field, the OBU randomly samples a fresh temporal nonce $u_d \in \mathbb{Z}_p^*$ that has no algebraic linkage to the actual chronological day index $d$. If $u_d$ were deterministically tied to the date, the insurer could easily sort the shuffled batch and reconstruct the driver's timeline. In addition, without a fresh random nonce, identical driving behaviors on different days would yield identical tags. This would expose daily repetitions and leave the system vulnerable to replay attacks. Therefore, combining $r_i$ and $u_d$ is strictly required to provide comprehensive spatial and temporal masking.
	
	The OBU constructs the daily polynomial authenticators (tags) for each feature $i$ as follows:
	\begin{equation}
		T_i^{(d)} = \alpha_t \cdot x_i^{(d)} + u_d \cdot r_i \pmod p
		\label{eq:mac_tag_daily}
	\end{equation}
	This specific structure (two scalar multiplications and one addition) is inspired by information-theoretic MACs. It is highly advantageous for homomorphic encryption systems because it avoids computationally expensive ciphertext-ciphertext multiplications while seamlessly distributing over the linear evaluation model. 
	
	Next, the OBU encrypts the driving data, the tags, and the temporal nonce under the insurer's public key $\mathit{pk}_{\text{Ins}}$. The daily tuple $\sigma^{(d)} = \big( \llbracket \mathbf{x}^{(d)} \rrbracket_{\mathit{pk}_{\text{Ins}}}, \llbracket \mathbf{T}^{(d)} \rrbracket_{\mathit{pk}_{\text{Ins}}}, \llbracket u_d \rrbracket_{\mathit{pk}_{\text{Ins}}} \big)$ is then transmitted to the TSP.
	
	Because the evaluation model $\mathcal{F}_{\text{Ins}}$ is linear, the TSP applies the identical homomorphic circuit to both the encrypted feature vector and the encrypted tag vector:
	\begin{equation}
		\llbracket \mathcal{R}_{\text{Ins}}^{(d)} \rrbracket_{\mathit{pk}_{\text{Ins}}} = \mathcal{F}_{\text{Ins}} \big( \llbracket \mathbf{x}^{(d)} \rrbracket_{\mathit{pk}_{\text{Ins}}} \big)
		\label{eq:mac_eval_risk}
	\end{equation}
	\begin{equation}
		\llbracket T_{\mathcal{R}}^{(d)} \rrbracket_{\mathit{pk}_{\text{Ins}}} = \mathcal{F}_{\text{Ins}} \big( \llbracket \mathbf{T}^{(d)} \rrbracket_{\mathit{pk}_{\text{Ins}}} \big)
		\label{eq:mac_eval_tag}
	\end{equation}
	At the end of epoch $t$, the TSP applies a random permutation $\pi$ to the $D$ evaluated tuples. This step generates a randomized batch $\mathcal{B} = \big\{ \big(\llbracket \mathcal{R}_{\text{Ins}}^{(\pi(j))} \rrbracket_{\mathit{pk}_{\text{Ins}}}, \llbracket T_{\mathcal{R}}^{(\pi(j))} \rrbracket_{\mathit{pk}_{\text{Ins}}}, \llbracket u_{\pi(j)} \rrbracket_{\mathit{pk}_{\text{Ins}}} \big) \big\}_{j=1}^D$, which is subsequently forwarded to the insurer.
	
	Upon decrypting tuple $j$, the insurer locally derives $\alpha_t$ and the base blinding vector $\mathbf{r}$. To guarantee that the TSP honestly executed the exact model with the correct coefficients, the insurer computes the expected base noise over its identical linear circuit just once per epoch: $r_{\text{total}} = \mathcal{F}_{\text{Ins}}(\mathbf{r}) \pmod p$. Since the TSP has no knowledge of the secret blinding vector $\mathbf{r}$, this calculation acts as an algebraic anchor. If the TSP alters the model or its coefficients, it cannot mathematically forge a tag that matches the insurer's hidden $r_{\text{total}}$. Execution integrity is definitively verified via the following algebraic congruence:
	\begin{equation}
		T_{\mathcal{R}}^{(j)} \overset{?}{\equiv} \alpha_t \cdot \mathcal{R}_{\text{Ins}}^{(j)} + u^{(j)} \cdot r_{\text{total}} \pmod p
		\label{eq:mac_verification}
	\end{equation}
	To prevent intra-epoch replay attacks (e.g., the TSP copying a favorable daily evaluation), the insurer asserts the uniqueness of $u^{(j)}$ within the batch. This approach ensures an $\mathcal{O}(1)$ verification complexity per tuple. It also guarantees strict temporal privacy, as the insurer exclusively observes random nonces $u^{(j)}$ with no algebraic linkage to the true temporal index $d$.

	\subsection{PROPOSED SCHEME}
	
	Before detailing the operational workflow of the proposed protocol, we briefly outline the system initialization phase and the distribution of cryptographic material. During the offline contract setup phase, the required asymmetric and symmetric keys are generated and provisioned to the participating entities based on their specific roles in the architecture. To ensure strict separation of duties and prevent unauthorized data access, each entity receives only the cryptographic keys necessary for its designated operations. Table~\ref{tab:key_management} summarizes the distribution of keys across OBU, RSU, TSP, and the insurer.
	
	\begin{table}[h!]
		\centering
		\caption{Cryptographic Key Distribution and Access Privilege Matrix Across System Entities}
		\label{tab:key_management}
		\footnotesize
		\renewcommand{\arraystretch}{1.2}
		\setlength{\tabcolsep}{4pt} 
		\begin{tabular}{@{} l c c c c c @{}}
			\toprule
			\textbf{Cryptographic Parameter} & \textbf{Notation} & \textbf{OBU} & \textbf{RSU} & \textbf{TSP} & \textbf{Ins.} \\
			\midrule
			\multicolumn{6}{@{}l}{\textit{\textbf{Asymmetric Encryption \& Decryption Layer}}} \\
			Risk Evaluation Public Key & $\mathit{pk}_{\text{Ins}}$ & $\bullet$ & $\circ$ & $\circ$ & $\circ$ \\
			Risk Evaluation Secret Key & $\mathit{sk}_{\text{Ins}}$ & $\circ$ & $\circ$ & $\circ$ & $\bullet$ \\
			Speed Verification Public Key & $\mathit{pk}_{\text{SV}}$ & $\bullet$ & $\circ$ & $\circ$ & $\circ$ \\
			Speed Verification Secret Key & $\mathit{sk}_{\text{SV}}$ & $\circ$ & $\circ$ & $\circ$ & $\bullet$ \\
			\midrule
			\multicolumn{6}{@{}l}{\textit{\textbf{Homomorphic Evaluation Layer}}} \\
			Risk Evaluation Key & $\mathit{evk}_{\text{Ins}}$ & $\circ$ & $\circ$ & $\bullet$ & $\circ$ \\
			Speed Verification Key & $\mathit{evk}_{\text{SV}}$ & $\circ$ & $\bullet$ & $\circ$ & $\circ$ \\
			\midrule
			\multicolumn{6}{@{}l}{\textit{\textbf{Authentication, Anonymity \& Routing Layer}}} \\
			Group Public Key & $\mathit{pk}_{\text{Group}}$ & $\bullet$ & $\bullet$ & $\bullet$ & $\circ$ \\
			OBU Group Signing Key & $\mathit{sk}_{\text{OBU}}$ & $\bullet$ & $\circ$ & $\circ$ & $\circ$ \\
			Group Manager Secret Key & $\mathit{sk}_{\text{Group}}$ & $\circ$ & $\circ$ & $\circ$ & $\bullet$ \\
			Symmetric HomMAC Key & $K$ & $\bullet$ & $\circ$ & $\circ$ & $\bullet$ \\
			Public Routing Identifier & $\mathit{ID}_{\text{Ins}}$ & $\bullet$ & $\bullet$ & $\bullet$ & $\bullet$ \\
			\bottomrule
		\end{tabular}
		
		\vspace{1ex}
		\raggedright \scriptsize \textit{Note:} $\bullet$ indicates key possession and authorization; $\circ$ indicates strict cryptographic isolation (no access).
	\end{table}
	
	\noindent\textbf{Step 1: Data Encapsulation and Homomorphic Tagging at the OBU.} 
	
	During the contract initialization phase, the cryptographic parameters (including the OBU's individual group signing key $\mathit{sk}_{\text{OBU}}$, the group public key $\mathit{pk}_{\text{Group}}$, the asymmetric encryption keys $\mathit{pk}_{\text{Ins}}$ and $\mathit{pk}_{\text{SV}}$, the symmetric MAC key $K$, and the public routing identifier $\mathit{ID}_{\text{Ins}}$) are provisioned within a tamper-resistant Hardware Security Module (HSM) or a Trusted Execution Environment (TEE) embedded in the OBU. The inclusion of $\mathit{ID}_{\text{Ins}}$ facilitates efficient ciphertext routing by intermediate edge nodes without exposing the policyholder's true identity (Algorithm~\ref{alg:obu_encapsulation}, line 1).
	
	During the operational phase, for each day $d$ within the billing epoch $t$, the OBU executes a sequence of numbered sub-steps as formalized in Algorithm~\ref{alg:obu_encapsulation}:
	
	\noindent\textit{Step 1.1 (Scalar Derivation):} The OBU derives the epoch-specific MAC base scalar $\alpha_t$ and the feature-specific blinding factors $r_i$ from the shared symmetric key $K$ using the cryptographic PRF (Algorithm~\ref{alg:obu_encapsulation}, lines 3--6).
	
	\noindent\textit{Step 1.2 (Tag Computation):} To support the verifiable execution integrity detailed in Section~\ref{subsec:hommac_temporal}, the OBU dynamically samples a daily temporal nonce $u_d \in \mathbb{Z}_p^*$ and computes the plaintext homomorphic authenticators (tags) $\mathbf{T}^{(d)}$ over the daily driving feature vector $\mathbf{x}^{(d)}$. Utilizing the epoch-unique $u_d$ ensures daily data freshness and facilitates deterministic malleability detection without requiring an auxiliary random nonce (Algorithm~\ref{alg:obu_encapsulation}, lines 8--11).
	
	\noindent\textit{Step 1.3 (Payload Structuring and Encryption):} In order to build the daily evaluation payload, the OBU first combines the policyholder's private identifier, the driving feature vector, the polynomial tags, and the temporal nonce into a unified plaintext tuple:
	\begin{equation}
		\mathcal{P}_{\text{raw}}^{(d)} = \big( \mathit{ID}, \, \mathbf{x}^{(d)}, \, \mathbf{T}^{(d)}, \, u_d \big)
		\label{eq:raw_payload_tuple}
	\end{equation}
	Because any untrusted edge infrastructure poses a major risk to the spatial and temporal anonymity of the driver, the OBU hides all elements of this package using homomorphic encryption. Specifically, it encrypts each component of the tuple $\mathcal{P}_{\text{raw}}^{(d)}$ individually under the insurer's public key $\mathit{pk}_{\text{Ins}}$. This computation yields the composite encrypted payload vector represented formally as $\llbracket \mathcal{P}^{(d)} \rrbracket_{\mathit{pk}_{\text{Ins}}}$ (Algorithm~\ref{alg:obu_encapsulation}, line 13).
	
	\noindent\textit{Step 1.4 (Speed Encryption):} Concurrently, the captured instantaneous speed parameter $v_{\text{sen}}^{(d)}$ receives a separate layer of encryption under the verification key $\mathit{pk}_{\text{SV}}$, producing the encrypted ciphertext $\llbracket v_{\text{sen}}^{(d)} \rrbracket_{\mathit{pk}_{\text{SV}}}$ for edge verification (Algorithm~\ref{alg:obu_encapsulation}, line 15).
	
	\noindent\textit{Step 1.5 (Anonymized Authentication):} Next, the OBU concatenates the encrypted payloads with the routing identifier $\mathit{ID}_{\text{Ins}}$ and applies a group signature protocol using its secret key $\mathit{sk}_{\text{OBU}}$ to generate the cryptographic signature $\sigma^{(d)}$. This signature provides a practical advantage: RSU can confirm the subscription status and validity of the vehicle without uncovering its true identity (Algorithm~\ref{alg:obu_encapsulation}, lines 17--18).
	
	\noindent\textit{Step 1.6 (Tuple Assembly and Storage):} Finally, the OBU merges these parts into the unified transmission tuple $\mathcal{C}^{(d)} = \big(\llbracket \mathcal{P}^{(d)} \rrbracket_{\mathit{pk}_{\text{Ins}}}, \, \llbracket v_{\text{sen}}^{(d)} \rrbracket_{\mathit{pk}_{\text{SV}}}, \, \mathit{ID}_{\text{Ins}}, \, \sigma^{(d)}\big)$ and broadcasts it to the edge infrastructure (Algorithm~\ref{alg:obu_encapsulation}, lines 20--21). To secure the device against physical theft, the OBU uses a local symmetric key to lock the raw telematics logs and random encryption values, storing them safely inside the TEE until the billing epoch concludes as a reliable backup for potential dispute resolution.
	
	\begin{algorithm}[H]
		\caption{OBU Cryptographic Payload Encapsulation \& Tagging}
		\label{alg:obu_encapsulation}
		\begin{algorithmic}[1]
			\renewcommand{\algorithmicrequire}{\textbf{Input:}}
			\renewcommand{\algorithmicensure}{\textbf{Output:}}
			
			\REQUIRE Daily feature vector $\mathbf{x}^{(d)}$ of size $n$, instantaneous speed $v_{\text{sen}}^{(d)}$, private $\mathit{ID}$, public $\mathit{ID}_{\text{Ins}}$, epoch $t$, prime modulus $p$.
			\ENSURE Unified daily cryptographic transmission tuple $\mathcal{C}^{(d)}$.
			
			\STATE \textit{// Retrieve parameters ($\mathit{pk}_{\text{Ins}}$, $\mathit{pk}_{\text{SV}}$, $\mathit{sk}_{\text{OBU}}$, $\mathit{pk}_{\text{Group}}$, $K$) from HSM/TEE}
			
			\STATE \textit{// Step 1.1: Derive epoch-specific and feature-specific MAC scalars}
			\STATE $\alpha_t \leftarrow F_K(t \parallel 0)$
			\FOR{each feature index $i \in \{1, \dots, n\}$}
			\STATE $r_i \leftarrow F_K(t \parallel i)$
			\ENDFOR
			
			\STATE \textit{// Step 1.2: Sample temporal nonce and compute daily HomMAC tags}
			\STATE $u_d \xleftarrow{\$} \mathbb{Z}_p^*$ \quad \textit{(sampled without replacement for epoch $t$)}
			\FOR{each feature index $i \in \{1, \dots, n\}$}
			\STATE $T_i^{(d)} \leftarrow \alpha_t \cdot x_i^{(d)} + u_d \cdot r_i \pmod p$
			\ENDFOR
			
			\STATE \textit{// Step 1.3: Homomorphic encryption of the insurance payload components}
			\STATE $\begin{aligned}
				\llbracket \mathcal{P}^{(d)} \rrbracket_{\mathit{pk}_{\text{Ins}}} \leftarrow \bigl( &\mathit{Enc}(\mathit{pk}_{\text{Ins}}, \mathit{ID}), \, \mathit{Enc}(\mathit{pk}_{\text{Ins}}, \mathbf{x}^{(d)}), \\[-0.5ex]
				&\mathit{Enc}(\mathit{pk}_{\text{Ins}}, \mathbf{T}^{(d)}), \, \mathit{Enc}(\mathit{pk}_{\text{Ins}}, u_d) \bigr)
			\end{aligned}$
			
			\STATE \textit{// Step 1.4: Encrypt instantaneous speed for ISAC verification}
			\STATE $\llbracket v_{\text{sen}}^{(d)} \rrbracket_{\mathit{pk}_{\text{SV}}} \leftarrow \mathit{Enc}\bigl(\mathit{pk}_{\text{SV}}, \, v_{\text{sen}}^{(d)}\bigr)$
			
			\STATE \textit{// Step 1.5: Generate group signature for anonymity and authentication}
			\STATE $msg \leftarrow \llbracket \mathcal{P}^{(d)} \rrbracket_{\mathit{pk}_{\text{Ins}}} \parallel \llbracket v_{\text{sen}}^{(d)} \rrbracket_{\mathit{pk}_{\text{SV}}} \parallel \mathit{ID}_{\text{Ins}}$
			\STATE $\sigma^{(d)} \leftarrow \mathit{Sign}\bigl(\mathit{sk}_{\text{OBU}}, \, \mathit{pk}_{\text{Group}}, \, msg \bigr)$
			
			\STATE \textit{// Step 1.6: Assemble the final transmission tuple}
			\STATE $\mathcal{C}^{(d)} \leftarrow \bigl(\llbracket \mathcal{P}^{(d)} \rrbracket_{\mathit{pk}_{\text{Ins}}}, \, \llbracket v_{\text{sen}}^{(d)} \rrbracket_{\mathit{pk}_{\text{SV}}}, \, \mathit{ID}_{\text{Ins}}, \, \sigma^{(d)}\bigr)$ 
			
			\RETURN $\mathcal{C}^{(d)}$
		\end{algorithmic}
	\end{algorithm}
	
	\noindent\textbf{Step 2: Message Authentication and ISAC-Assisted Speed Verification at the RSU.} 
	
	As detailed in Algorithm~\ref{alg:speed_rsu}, upon receiving the uplink electromagnetic signal carrying the encapsulated payload $\mathcal{C}^{(d)}$ from the OBU, the RSU concurrently decodes the communication data and extracts the physical sensing parameters. The verification procedure is structured into three sequential sub-steps:
	
	\noindent\textit{Step 2.1 (Group Signature Authentication):} Initially, the RSU performs cryptographic authentication utilizing the group public key ($\mathit{pk}_{\text{Group}}$) to validate the data origin and integrity of the message. If the group signature verification fails, the payload is immediately rejected and discarded to prevent computational resource exhaustion at the edge infrastructure (Algorithm~\ref{alg:speed_rsu}, lines 2--6).
	
	\noindent\textit{Step 2.2 (Reliability-Aware Speed Verification):} Conversely, upon successful authentication, the RSU proceeds to the ISAC-assisted speed verification phase, as formalized in Section~\ref{subsec:velocity_verification}. The RSU computes the true physical speed ($v_{\text{isac}}$) from the extracted Doppler shift and evaluates the channel SNR  ($\gamma$). To prevent false penalization due to degraded channel conditions, the RSU compares $\gamma$ against the predefined reliability threshold $\gamma_{\text{th}}$. If $\gamma \ge \gamma_{\text{th}}$, the RSU executes the lightweight verification circuit $\mathcal{F}_{\text{SV}}$ utilizing the evaluation key $\mathit{evk}_{\text{SV}}$. This operation performs a mixed-domain homomorphic subtraction between the encrypted self-reported speed ($\llbracket v_{\text{sen}}^{(d)} \rrbracket_{\mathit{pk}_{\text{SV}}}$) and the plaintext physical measurement ($v_{\text{isac}}$). Alternatively, if the channel threshold is not met ($\gamma < \gamma_{\text{th}}$), a deterministic $\text{FLAG}_{\text{uncert}}$ is generated to indicate sensing uncertainty (Algorithm~\ref{alg:speed_rsu}, lines 8--14).
	
	\noindent\textit{Step 2.3 (Message Assembly and Forwarding):} Finally, the verification result, denoted formally as $\mathcal{R}_{\text{SV}}^{(d)}$, is assembled alongside the original vehicular payload $\mathcal{C}^{(d)}$ into a forwarding message $\mathit{RSU}_M$ and transmitted to the TSP (Algorithm~\ref{alg:speed_rsu}, lines 16--17).
	
	\begin{algorithm}[H]
		\caption{ISAC-Assisted Homomorphic Speed Verification at RSU}
		\label{alg:speed_rsu} 
		\begin{algorithmic}[1]
			\renewcommand{\algorithmicrequire}{\textbf{Input:}}
			\renewcommand{\algorithmicensure}{\textbf{Output:}}
			
			\REQUIRE Vehicular payload $\mathcal{C}^{(d)} = \bigl(\llbracket \mathcal{P}^{(d)} \rrbracket_{\mathit{pk}_{\text{Ins}}}, \, \llbracket v_{\text{sen}}^{(d)} \rrbracket_{\mathit{pk}_{\text{SV}}}, \, \mathit{ID}_{\text{Ins}}, \, \sigma^{(d)} \bigr)$, physical speed $v_{\text{isac}}$, channel SNR $\gamma$, reliability threshold $\gamma_{\text{th}}$, evaluation key $\mathit{evk}_{\text{SV}}$.
			\ENSURE Forwarding message $\mathit{RSU}_M$ containing the verification result.
			
			\STATE \textit{// Step 2.1: Cryptographically verify the group signature}
			\STATE $msg \leftarrow \llbracket \mathcal{P}^{(d)} \rrbracket_{\mathit{pk}_{\text{Ins}}} \parallel \llbracket v_{\text{sen}}^{(d)} \rrbracket_{\mathit{pk}_{\text{SV}}} \parallel \mathit{ID}_{\text{Ins}}$
			\IF{$\mathit{Verify}(\sigma^{(d)}, \, msg, \, \mathit{pk}_{\text{Group}}) == \text{False}$}
			\STATE \textbf{reject} Invalid payload; \textbf{abort} protocol
			\RETURN $\perp$
			\ENDIF
			
			\STATE \textit{// Step 2.2: Evaluate ISAC physical-layer reliability}
			\IF{$\gamma \ge \gamma_{\text{th}}$}
			\STATE \textit{// Execute mixed-domain homomorphic subtraction}
			\STATE $\mathcal{R}_{\text{SV}}^{(d)} \leftarrow \llbracket v_{\text{sen}}^{(d)} \rrbracket_{\mathit{pk}_{\text{SV}}} \ominus v_{\text{isac}}$
			\ELSE
			\STATE \textit{// Flag uncertainty due to severe multipath/fading}
			\STATE $\mathcal{R}_{\text{SV}}^{(d)} \leftarrow \text{FLAG}_{\text{uncert}}$
			\ENDIF
			
			\STATE \textit{// Step 2.3: Assemble payload for TSP forwarding}
			\STATE $\mathit{RSU}_M \leftarrow (\mathcal{R}_{\text{SV}}^{(d)}, \, \mathcal{C}^{(d)})$
			\RETURN $\mathit{RSU}_M$
		\end{algorithmic}
	\end{algorithm}
	
	\noindent\textbf{Step 3: Message Authentication and Risk Evaluation Circuit Execution at the TSP.} 
	
	Upon receiving the forwarded message $\mathit{RSU}_M^{(d)}$ from the RSU, the TSP executes the computational and obfuscation workflow formalized in Algorithm~\ref{alg:tsp_risk_calculation} through five distinct sub-steps:
	
	\noindent\textit{Step 3.1 (Signature Re-verification):} The TSP first performs an end-to-end cryptographic verification of the group signature utilizing the corresponding group public key ($\mathit{pk}_{\text{Group}}$) to ensure data integrity across the routing hops. If the verification fails, the record is discarded (Algorithm~\ref{alg:tsp_risk_calculation}, lines 3--7).
	
	\noindent\textit{Step 3.2 (Payload Unpacking):} Upon successful authentication, the TSP accesses and extracts the elements of the OBU's encrypted payload vector, specifically isolating $\llbracket \mathbf{x}^{(d)} \rrbracket_{\mathit{pk}_{\text{Ins}}}$, $\llbracket \mathbf{T}^{(d)} \rrbracket_{\mathit{pk}_{\text{Ins}}}$, and $\llbracket u_d \rrbracket_{\mathit{pk}_{\text{Ins}}}$ from $\llbracket \mathcal{P}^{(d)} \rrbracket_{\mathit{pk}_{\text{Ins}}}$ (Algorithm~\ref{alg:tsp_risk_calculation}, line 9).
	
	\noindent\textit{Step 3.3 (Homomorphic Circuit Execution):} To compute the final driving risk factor while maintaining verifiable execution integrity (as formalized in Section~\ref{subsec:hommac_temporal}), the TSP executes the insurer's risk evaluation circuit $\mathcal{F}_{\text{Ins}}$ utilizing the dedicated evaluation key $\mathit{evk}_{\text{Ins}}$. This homomorphic circuit is applied concurrently to both the encrypted feature vector and the encrypted tag vector, yielding the encrypted daily risk factor $\llbracket \mathcal{R}_{\text{Ins}}^{(d)} \rrbracket_{\mathit{pk}_{\text{Ins}}}$ and its aggregated tag $\llbracket T_{\mathcal{R}}^{(d)} \rrbracket_{\mathit{pk}_{\text{Ins}}}$ (Algorithm~\ref{alg:tsp_risk_calculation}, lines 11--12).
	
	\noindent\textit{Step 3.4 (Local Repository Storage):} To mitigate temporal profiling, the TSP does not immediately transmit these daily results to the insurer. Instead, the evaluated tuple (comprising the computed risk score, the aggregated tag, the encrypted temporal nonce, and the ISAC verification result $\mathcal{R}_{\text{SV}}^{(d)}$) is stored in the TSP's local repository (Algorithm~\ref{alg:tsp_risk_calculation}, line 14).
	
	\noindent\textit{Step 3.5 (Batch Shuffling and Transmission):} To address temporal inference attacks (a privacy vulnerability where the sequential transmission of daily results allows the insurer to reconstruct chronological behavioral routines), the proposed protocol delays data transmission until the billing epoch concludes. Let $D$ denote the total number of accumulated daily records. At the end of the epoch, the TSP applies a random permutation $\pi$ over the indices $\{1, \dots, D\}$. This shuffling mechanism breaks the chronological linkage, producing a randomized batch $\mathcal{B}$ that prevents time-series alignment analysis by the insurer (Algorithm~\ref{alg:tsp_risk_calculation}, lines 17--19).
	
	\begin{algorithm}[H]
		\caption{Homomorphic Risk Evaluation and TSP Batch Generation}
		\label{alg:tsp_risk_calculation}
		\begin{algorithmic}[1]
			\renewcommand{\algorithmicrequire}{\textbf{Input:}}
			\renewcommand{\algorithmicensure}{\textbf{Output:}}
			
			\REQUIRE Forwarded RSU payloads $\mathit{RSU}_M^{(d)} = (\mathcal{R}_{\text{SV}}^{(d)}, \, \mathcal{C}^{(d)})$ for each day $d \in \{1, \dots, D\}$ in epoch $t$.
			\ENSURE A chronologically shuffled evaluation batch $\mathcal{B}$ for the insurer.
			
			\FOR{each day $d \in \{1, \dots, D\}$}
			\STATE \textit{// Step 3.1: End-to-end group signature verification}
			\STATE $msg \leftarrow \llbracket \mathcal{P}^{(d)} \rrbracket_{\mathit{pk}_{\text{Ins}}} \parallel \llbracket v_{\text{sen}}^{(d)} \rrbracket_{\mathit{pk}_{\text{SV}}} \parallel \mathit{ID}_{\text{Ins}}$
			\IF{$\mathit{Verify}(\sigma^{(d)}, \, msg, \, \mathit{pk}_{\text{Group}}) == \text{False}$}
			\STATE \textbf{reject} Invalid payload; \textbf{discard} daily record
			\STATE \textbf{continue} to next iteration
			\ENDIF
			
			\STATE \textit{// Step 3.2: Access encrypted payload components}
			\STATE Retrieve $\llbracket \mathbf{x}^{(d)} \rrbracket_{\mathit{pk}_{\text{Ins}}}$, $\llbracket \mathbf{T}^{(d)} \rrbracket_{\mathit{pk}_{\text{Ins}}}$, and $\llbracket u_d \rrbracket_{\mathit{pk}_{\text{Ins}}}$ from the vector $\llbracket \mathcal{P}^{(d)} \rrbracket_{\mathit{pk}_{\text{Ins}}}$
			
			\STATE \textit{// Step 3.3: Execute linear risk evaluation circuit and HomMAC tag aggregation}
			\STATE $\llbracket \mathcal{R}_{\text{Ins}}^{(d)} \rrbracket_{\mathit{pk}_{\text{Ins}}} \leftarrow \mathcal{F}_{\text{Ins}}\bigl(\llbracket \mathbf{x}^{(d)} \rrbracket_{\mathit{pk}_{\text{Ins}}}, \, \mathit{evk}_{\text{Ins}}\bigr)$
			\STATE $\llbracket T_{\mathcal{R}}^{(d)} \rrbracket_{\mathit{pk}_{\text{Ins}}} \leftarrow \mathcal{F}_{\text{Ins}}\bigl(\llbracket \mathbf{T}^{(d)} \rrbracket_{\mathit{pk}_{\text{Ins}}}, \, \mathit{evk}_{\text{Ins}}\bigr)$
			
			\STATE \textit{// Step 3.4: Store daily evaluated tuple in the local repository}
			\STATE $\mathit{TSP}^{(d)} \leftarrow \bigl(\llbracket \mathcal{R}_{\text{Ins}}^{(d)} \rrbracket_{\mathit{pk}_{\text{Ins}}}, \, \llbracket T_{\mathcal{R}}^{(d)} \rrbracket_{\mathit{pk}_{\text{Ins}}}, \, \llbracket u_d \rrbracket_{\mathit{pk}_{\text{Ins}}}, \, \mathcal{R}_{\text{SV}}^{(d)} \bigr)$
			\ENDFOR
			
			\STATE \textit{// Step 3.5: At the conclusion of epoch $t$, apply random permutation $\pi$}
			\STATE Generate permutation $\pi$ over indices $\{1, \dots, D\}$
			\STATE $\mathcal{B} \leftarrow \big\{ \mathit{TSP}^{(\pi(j))} \big\}_{j=1}^D$
			
			\RETURN $\mathcal{B}$
		\end{algorithmic}
	\end{algorithm}
	
	\noindent\textbf{Step 4: Result Decryption, Integrity Verification, and Final Extraction by the Insurer.} 
	
	In the final phase of the billing epoch, the insurer receives the chronologically shuffled batch $\mathcal{B}$ from the TSP. Utilizing its exclusive secret keys ($\mathit{sk}_{\text{Ins}}$ and $\mathit{sk}_{\text{SV}}$), the insurer decrypts the elements of each tuple $j \in \{1, \dots, D\}$. Specifically, $\mathit{sk}_{\text{Ins}}$ is employed to reveal the private policyholder identifier ($\mathit{ID}$), the homomorphically evaluated risk factor ($\mathcal{R}_{\text{Ins}}^{(j)}$), the aggregated HomMAC tag ($T_{\mathcal{R}}^{(j)}$), and the temporal nonce ($u^{(j)}$). Concurrently, the insurer utilizes $\mathit{sk}_{\text{SV}}$ to decrypt the ISAC-assisted speed verification result ($\mathcal{R}_{\text{SV}}^{(j)}$). 
	
	Before utilizing the risk factor for premium calculation, the insurer performs two verification checks. First, it verifies the execution integrity of the TSP's homomorphic aggregation via the algebraic congruence defined in Eq.~\eqref{eq:mac_verification}. Second, the insurer validates the physical speed integrity by confirming that the absolute error $\mathcal{R}_{\text{SV}}^{(j)}$ falls within the predefined physical tolerance threshold $\epsilon$, as stipulated in Eq.~\eqref{eq:threshold_check}. Tuples successfully passing both validations are aggregated for the final premium adjustment, while those failing are flagged for potential spoofing or computation tampering.
	
	\noindent\textbf{Dispute Resolution Mechanism:} 
	The group signature $\sigma^{(d)}$ securely links all of the daily payload together. Thanks to this cryptographic bond, no single party can deny the transmission of the data. As a result, the system establishes a clear and undeniable trail for any future audits. Every now and then, a dispute might arise between the two parties.  In the event of such a conflict, the policyholder can extract the locally encrypted logs directly from the OBU. The driver decrypts these specific files on a local level first. After that, the driver securely reveals the raw driving data along with the random values used for the initial encryption. With all of this information, the logs are simply re-encrypted from scratch. This step mathematically checks if the new results match the original ciphertexts exactly. Such a deterministic verification offers a definitive cryptographic proof of the exact inputs from the driver. Ultimately, this method creates a fair and auditable framework for the resolution of conflicts. Above all, it ensures that no raw driving data is ever left unprotected inside the vehicle.

	\section{PRIVACY AND SECURITY ANALYSIS}
	
	This section evaluates the proposed \textit{SP2UBI} framework in three phases. First, an informal analysis verifies how the architecture satisfies core security and privacy requirements. Second, a formal evaluation quantifies the privacy and confidentiality guarantees of the protocol. Finally, a comparative analysis benchmarks the scheme against existing vehicular telematics models.
	
	\subsection{Evaluation of Privacy and Security Requirements}
	This subsection examines how foundational privacy and security requirements are fulfilled within the \textit{SP2UBI} scheme. 
	
	\emph{Unlinkability:} Unlinkability is established by implementing a dual-layered solution. On the edge side, the identity of the policyholder is concealed from the RSU by using a group signature, while the replacement of the private identifier with the public routing identifier ($\mathit{ID}_{\text{Ins}}$) ensures unlinkability at the edge node. On the cloud side, epoch-end shuffling by the TSP mathematically disrupts temporal correlations between ciphertexts and ensures that daily risk factors cannot be chronologically linked to spatiotemporal trajectories. 
	
	\emph{Non-repudiation:} This requirement is met cryptographically due to the creation of an unforgeable group signature that authenticates the data origin and prevents the policyholder from denying message transmission. In addition, the secure archiving of deterministic plaintext logs inside the tamper-resistant OBU ensures cryptographic evidence to resolve any further billing disputes. 
	
	\emph{Confidentiality and Non-disclosure:} Confidentiality is provided at all infrastructure levels. The protocol intentionally uses coarse-grained statistical aggregation and does not transfer fine-grained GPS trajectories. Specifically, all risk and speed assessments are performed within the homomorphically encrypted domain. Thus, intermediary entities (e.g., the TSP) become blind evaluators of the traffic data, which ensures that raw vehicular telemetry remains inaccessible outside the trusted OBU-insurer perimeter. 
	
	\emph{Unobservability:} Data transmission becomes unobservable due to decoupling the data from its real-time spatial context. The TSP consistently removes any physical RSU identifiers and geographic metadata from the transmitted payloads to forward an aggregated, non-sequential batch to the insurer. As a result, time-series analysis of the transmitted batches will not allow the insurer to infer the daily commuting patterns of the policyholder. 
	
	\emph{Data Sovereignty (Content Awareness):} Since the initial telematics aggregation and cryptographic encapsulation are performed solely within the user-controlled OBU, data sovereignty is preserved. This gives the policyholder full content awareness before any transmission of data to external parties. 
	
	\emph{Insurer's Evaluation Privacy:} The confidentiality of the insurer's proprietary risk assessment model is protected. Due to the conversion of the risk model into a corresponding FHE circuit, the TSP is able to perform all computations without having plaintext access to potentially sensitive model weights and telematics data inputs. 
	
	\emph{Execution Integrity and Verifiability:} Unlike vulnerable systems susceptible to silent computational failures or malicious cloud tampering, \textit{SP2UBI} guarantees execution integrity without sacrificing data confidentiality. The implementation of the decoupled HomMAC protocol allows the insurance provider to mathematically verify that the TSP correctly executes the intended linear circuit ($\mathcal{F}_{\text{Ins}}$). Furthermore, temporal nonces protect against intra-epoch replay attacks and intra-batch ciphertext substitution.
	
	\subsection{Formal Evaluation of Technical Privacy Metrics}
	To rigorously quantify the privacy guarantees of the protocol against a rational adversary, the structured technical metrics proposed in the systematic survey by Wagner and Eckhoff \cite{Wagner2015TechnicalPM} are adapted. Given the context of UBI, the primary goal is to prevent behavioral profiling and pattern inference. Accordingly, the proposed scheme is evaluated based on the formal criteria of day-to-day unlinkability and informational entropy.
	
	\subsubsection{Mitigation of Temporal and Behavioral Profiling}
	Over a billing epoch of $D$ days (e.g., $D = 30$), the protocol generates a set of daily risk assessment vectors and speed verification results. Although the insurer identifies the policyholder for the final premium calculation, the proposed protocol prevents the insurer from linking any specific daily evaluation to its exact chronological calendar date. This privacy bound is established under three core architectural assumptions: (i) the insurer only accesses the authorized identity and the unordered batch of decrypted daily results $\mathcal{B}$, without external side-channels; (ii) prior to transmission, the TSP strictly shuffles the accumulated daily records using a cryptographically secure pseudo-random permutation uniformly selected from the symmetric group of all possible arrangements $\Pi$ ($|\Pi| = D!$); and (iii) the decrypted parameters, including the HomMAC temporal nonces $u^{(j)}$, are structurally independent and contain no deterministic temporal identifiers linking back to the true day index $d$.
	
	Following \cite{Wagner2015TechnicalPM}, the degree of daily unlinkability ($\text{priv}_{\text{DUE}}$) is quantified via the Shannon entropy of the adversary's probability distribution $p(\pi)$ over the hypothesis space $\Pi$:
	\begin{equation}
		\text{priv}_{\text{DUE}} \equiv H(\Pi) = -\sum_{\pi \in \Pi} p(\pi) \log_2 p(\pi)
		\label{eq:shannon_entropy}
	\end{equation}
	
	\noindent\textbf{Theorem 1.} \textit{The proposed epoch-end shuffling mechanism satisfies the day-to-day unlinkability criterion ($\text{priv}_{\text{DUE}}$), bounding the adversarial insurer's probability of chronologically reconstructing a policyholder's behavioral profile over a $D$-day billing epoch to $\frac{1}{D!}$.}
	
	\begin{IEEEproof}
		The daily unlinkability metric is evaluated across two distinct operational states:
		
		\textit{Case 1 (Baseline State Without Shuffling):} If the TSP transmits daily evaluations chronologically, the sequence is deterministic, yielding $p(\pi_{\text{true}}) = 1$ and $p(\pi \neq \pi_{\text{true}}) = 0$. Substituting these values into Eq.~\eqref{eq:shannon_entropy} results in $H(\Pi) = 0$, indicating deterministic temporal linkability and enabling behavioral trajectory reconstruction by the insurer.
		
		\textit{Case 2 (Protected State With Uniform Shuffling):} Due to the cryptographically uniform permutation applied by the TSP and the absence of side-channels, the adversarial insurer's prior probability over the hypothesis space $\Pi$ is strictly uniform: $p(\pi) = 1/D!, \, \forall \pi \in \Pi$. Substituting this uniform distribution into Eq.~\eqref{eq:shannon_entropy} yields the upper bound for informational entropy under a uniform distribution:
		\begin{equation}
			H(\Pi) = -\sum_{\pi \in \Pi} \frac{1}{D!} \log_2 \left(\frac{1}{D!}\right) = \log_2(D!)
			\label{eq:maximal_entropy}
		\end{equation}
		For a standard monthly billing epoch where $D = 30$, Eq.~\eqref{eq:maximal_entropy} yields $H(\Pi) = \log_2(30!) \approx 107.74$ bits. Consequently, the brute-force success probability for an adversarial insurer attempting to deduce the exact chronological configuration of a policyholder's behavioral profile is bounded by $2^{-107.74}$, which is statistically negligible, thereby mitigating day-to-day linkability risks.
	\end{IEEEproof}
	
	\subsubsection{Anonymity Against External Attackers, RSUs, and the TSP}
	In VANET environments, the primary tracking threat stems from the linkability of transmitted messages. To mitigate this vulnerability, the proposed framework leverages $k$-anonymity to mask the sender's identity within an equivalence class. Following the Wagner-Eckhoff taxonomy \cite{Wagner2015TechnicalPM}, an equivalence class $E_i$ represents the set of all potential users who could have generated a specific transmission from the adversary's perspective. The system satisfies $k$-anonymity ($\text{priv}_{\text{KA}}$) if and only if the cardinality of every equivalence class across $X$ independent channels boundedly meets or exceeds $k$:
	\begin{equation}
		\text{priv}_{\text{KA}} \ge k \iff \forall i \in \{1,\dots,X\}: |E_i| \geq k
		\label{eq:k-anno}
	\end{equation}
	
	\noindent\textbf{Theorem 2.} \textit{The proposed packet encapsulation protocol satisfies $k$-anonymity for the vehicle against honest-but-curious RSUs, the TSP, and external eavesdroppers.}
	
	\begin{IEEEproof}
		Let $E$ be an arbitrary equivalence class of users associated with a captured transmission payload $\mathcal{C}^{(d)}$. According to the operational design, every validated packet contains a group signature $\sigma^{(d)}$ verifiable solely via the group public key $\mathit{pk}_{\text{Group}}$ assigned to a distinct membership set $G$. Due to the indistinguishability property of the underlying group signature scheme, the cryptographic signatures leak no identifying information beyond group membership. Consequently, an adversary (whether an external attacker or the backend TSP) can only deduce that the packet originated from some authorized member within the group $G$. It follows that the adversary's hypothesis space is restricted to the group boundaries, establishing an exact structural equivalence between the anonymity class and the group population, denoted as $E = G$. Given that the initialization phase provisions each cryptographic group with $k$ distinct policyholders ($|G| = k$), the cardinality of the equivalence class is deterministically bounded by:
		\begin{equation}
			|E| = |G| = k
		\end{equation}
		Substituting this result into Eq.~\eqref{eq:k-anno} confirms that $\forall i: |E_i| = k \geq k$, thereby formally satisfying the $k$-anonymity metric.
	\end{IEEEproof}
	
	\subsubsection{Semantic Security Against Intermediary Entities}
	To safeguard raw telematics streams against honest-but-curious intermediaries, the framework enforces semantic security throughout the data lifecycle. All data vectors generated by the OBU are encrypted prior to transmission using an FHE scheme that satisfies indistinguishability under chosen-plaintext attacks ($\mathit{IND\text{-}CPA}$). Under this paradigm, the RSU and TSP are provisioned exclusively with evaluation keys ($\mathit{evk}_{\text{SV}}$ and $\mathit{evk}_{\text{Ins}}$, respectively), while the corresponding decryption keys remain strictly private to the insurer. 
	
	Following the Wagner-Eckhoff classification \cite{Wagner2015TechnicalPM}, the degree of confidentiality via a cryptographic game ($\text{priv}_{\text{CG}}$) is formulated as a binary success indicator:
	\begin{equation}
		\text{priv}_{\text{CG}} \equiv \begin{cases} 
			1 & \text{if } \Pr[\mathcal{A}(c_b) = b] \leq \frac{1}{2} + \epsilon(\kappa) \\ 
			0 & \text{otherwise} 
		\end{cases}
		\label{eq:CG1}
	\end{equation}
	where $\mathcal{A}$ denotes a polynomial-time adversary, $c_b = \mathsf{Enc}(m_b, \mathit{pk})$ is the challenge ciphertext generated from two chosen plaintexts $m_0, m_1$ of equal length, and $\epsilon(\kappa)$ is a negligible function of the security parameter $\kappa$. 
	
	\noindent\textbf{Theorem 3.} \textit{The proposed framework satisfies semantic security ($\mathit{IND\text{-}CPA}$) for all vehicular telematics payloads against honest-but-curious intermediary entities, ensuring $\text{priv}_{\text{CG}} = 1$.}
	
	\begin{IEEEproof}
		Consider an adversarial intermediary (e.g., the TSP) executing the linear risk evaluation circuit $\mathcal{F}_{\text{Ins}}$. Since the underlying FHE construction is established to be $\mathit{IND\text{-}CPA}$ secure, no computationally efficient algorithm can distinguish between the ciphertexts of distinct plaintexts. Consequently, the advantage of $\mathcal{A}$ over a random guess is bounded by $\epsilon(\kappa)$, limiting its winning probability to $\frac{1}{2} + \epsilon(\kappa)$. This satisfies the upper branch of Eq.~\eqref{eq:CG1}, yielding $\text{priv}_{\text{CG}} = 1$. Thus, even with full exposure to the evaluation keys and intermediate execution states, the telematics data remains semantically secure and computationally inaccessible to unauthorized entities.
	\end{IEEEproof}
	
	\subsubsection{Source Location Obfuscation and Adversarial Success Probability}
	To counter localized trajectory tracking, the insurer is modeled as an active adversary attempting to infer the policyholder's spatiotemporal distribution via the identity of the forwarding  RSU. Let $\mathcal{X}$ denote the universe of all deployed RSUs, and $X \in \mathcal{X}$ be a random variable representing the actual source RSU. Let $Y \in \mathcal{Y}$ be the random variable representing the observable metadata intercepted by the insurer. Upon observing $Y=y$, the adversary executes an optimal estimation function $\hat{X}: \mathcal{Y} \to \mathcal{X}$ to guess the true source. Following \cite{Wagner2015TechnicalPM}, the global adversarial success probability $P_{\text{succ}}$ is bounded by the maximum posterior probability:
	\begin{equation}
		P_{\text{succ}}(y) = \max_{i \in \mathcal{X}} \Pr(X = i \mid Y = y)
		\label{eq:Psuc_def_new}
	\end{equation}
	
	\noindent\textbf{Theorem 4.} \textit{The TSP-mediated architecture reduces the adversarial success probability for source location tracking from a deterministic baseline ($P_{\text{succ}}^{\mathrm{base}} = 1$) to a probabilistic bound ($P_{\text{succ}}^{\mathrm{proposed}} = \frac{1}{|S|}$), where $|S|$ is the cardinality of the regional RSU anonymity set.}
	
	\begin{IEEEproof}
		The adversarial tracking capabilities across the two architectural paradigms are evaluated as follows:
		
		\textit{Case 1 (Baseline Direct Channel):} In an architecture without a mediating proxy, each RSU forwards packets directly to the insurer along with its explicit identity. Thus, the observed metadata contains a deterministic mapping where $Y = i$. The posterior probability reduces to a degenerate distribution, modeled mathematically via the Kronecker delta:
		\begin{equation}
			\Pr(X = i \mid Y = y) = \delta_{iy} = \begin{cases} 1 & \text{if } i = y \\ 0 & \text{otherwise} \end{cases}
			\label{eq:posterior_base}
		\end{equation}
		Substituting Eq.~\eqref{eq:posterior_base} into Eq.~\eqref{eq:Psuc_def_new} yields $P_{\text{succ}}^{\mathrm{base}} = 1$, demonstrating that the insurer can identify the exact source location deterministically.
		
		\textit{Case 2 (Proposed TSP-Mediated Obfuscation):} To prevent tracking via edge node identifiers, the proposed protocol utilizes the TSP as a privacy-preserving proxy. The TSP actively intercepts RSU transmissions, stripping all localized geographic markers and physical RSU identifiers before batching and delivering the statistical payload to the insurer. Consequently, the insurer can only map the metadata $y$ to an anonymity set of compatible RSUs covering the region over the $D$-day epoch, denoted as $S \subseteq \mathcal{X}$. Assuming no side-channel bias, the posterior distribution over $S$ is strictly uniform:
		\begin{equation}
			\Pr(X = i \mid Y = y) = \begin{cases} \frac{1}{|S|} & \text{if } i \in S \\ 0 & \text{otherwise} \end{cases}
			\label{eq:posterior_uniform_new}
		\end{equation}
		Under this protected state, the adversary's optimal guessing strategy yields a significantly mitigated success bound:
		\begin{equation}
			P_{\text{succ}}^{\mathrm{proposed}} = \max_{i \in S} \left( \frac{1}{|S|} \right) = \frac{1}{|S|}
			\label{eq:Psuc_proposed_new}
		\end{equation}
		
		Assuming a non-trivial regional deployment where $|S| \geq 2$, comparing the two derivations demonstrates a strict mathematical reduction in adversarial tracking capability:
		\begin{equation}
			P_{\text{succ}}^{\mathrm{proposed}} = \frac{1}{|S|} < P_{\text{succ}}^{\mathrm{base}} = 1
			\label{eq:Psuc_compare}
		\end{equation}
		This formalizes that the TSP-mediated architecture enhances source location obfuscation, preventing reliance on vulnerable edge nodes and forcing the insurer's tracking capability to scale inversely with the size of the spatial anonymity set.
	\end{IEEEproof}
	
	\subsubsection{Collusion Analysis Among Entities}
	The potential collusion scenarios among participating entities, along with their respective adversarial targets and the corresponding mitigation mechanisms, are systematically summarized in Table~\ref{tab:collusion}. By enforcing coarse-grained statistical data encapsulation, evaluating homomorphic operations exclusively over these aggregated daily states, and leveraging cryptographic shuffling, \textit{SP2UBI} preserves user privacy and execution integrity against multi-entity collusion scenarios.

	\begin{table}[htbp]
		
		\centering
		
		\caption{Collusion Scenarios and Mitigation Mechanisms}
		
		\label{tab:collusion}
		
		\renewcommand{\arraystretch}{1.3}
		
		\footnotesize 
		
		\begin{tabularx}{\linewidth}{>{\raggedright\arraybackslash}p{0.35\linewidth} >{\raggedright\arraybackslash}X}
			
			\toprule
			
			\textbf{Collusion Scenario \& Target} & \textbf{Mitigation Mechanism} \\
			
			\midrule
			
			\textbf{OBU $\leftrightarrow$ \{RSU, TSP\}} \newline
			
			\textit{Target:} Reverse-engineering proprietary risk models ($\mathcal{F}_{\text{Ins}}, \mathcal{F}_{\text{SV}}$) or manipulating evaluation circuits. & 
			
			The intermediaries are provisioned exclusively with evaluation keys ($\mathit{evk}$), ensuring the semantic security ($\mathit{IND\text{-}CPA}$) of the underlying FHE scheme. Furthermore, any malicious deviation from the designated circuits is deterministically detected by the insurer via the decoupled HomMAC algebraic verification. \\
			
			\midrule
			
			\textbf{Insurer $\leftrightarrow$ \{RSU, TSP\}} \newline
			
			\textit{Target:} Extraction of fine-grained spatiotemporal trajectories or chronological profiling. & 
			
			The protocol enforces coarse-grained statistical aggregation natively at the OBU to prevent fine-grained trajectory reconstruction. Consequently, the colluding entities only access coarse-grained data without spatio-temporal metadata, thereby preserving a level of privacy. \\

			\bottomrule
			
		\end{tabularx}
		
	\end{table}

	\begin{table*}[h]
		\centering
		\caption{Comparative Analysis of the Proposed \textit{SP2UBI} Framework with State-of-the-Art UBI Schemes}
		\label{tab:comparative_analysis}
		\renewcommand{\arraystretch}{1.4}
		\resizebox{\textwidth}{!}{
			\begin{tabular}{l l c c c c c}
				\toprule
				\multirow{2}{*}{\textbf{Framework / Scheme}} & \multirow{2}{*}{\textbf{Primary Cryptographic / Technical Mechanism}} & \multicolumn{5}{c}{\textbf{Security and Privacy Capabilities}} \\
				\cmidrule(lr){3-7}
				& & \textbf{E2E Ciphertext} & \textbf{Physical-Layer} & \textbf{Spatiotemporal} & \textbf{Conditional} & \textbf{Confidential} \\
				& & \textbf{Confidentiality\textsuperscript{a}} & \textbf{Fraud Resistance\textsuperscript{b}} & \textbf{Unlinkability} & \textbf{Anonymity} & \textbf{Anomaly Detection\textsuperscript{c}} \\
				\midrule
				\textbf{PriPAYD} \cite{Troncoso2007PriPAYDPP} & Local Processing + Tamper-resistant HW & \ding{55} & $\sim$ & \ding{51} & \ding{55} & \ding{55} \\
				\textbf{VPriv} \cite{Popa2009VPrivPP} & ZKPs + Cryptographic Commitments & \ding{51} & $\sim$ & $\sim$ & \ding{51} & $\sim$ \\
				\textbf{PRIDE} \cite{Wan2018PRIDEAP} & Matrix-based Homomorphic Encryption & \ding{55} & \ding{55} & \ding{55} & \ding{55} & \ding{55} \\
				\textbf{DUBI} \cite{Qi2021ScalableDP} & Pedersen Commitments + NIZKPs & \ding{51} & $\sim$ & $\sim$ & \ding{51} & $\sim$ \\
				\textbf{Huang et al.} \cite{Huang2023BlockchainAssistedPC} & Paillier Cryptosystem + ZKPs & \ding{51} & \ding{55} & \ding{51} & \ding{51} & $\sim$ \\
				\textbf{BE-VIP} \cite{Sahu2024BlockchainAM} & Local ML (Logistic Regression) + IPFS & \ding{55} & \ding{55} & \ding{51} & \ding{51} & $\sim$ \\
				\midrule
				\rowcolor{gray!15}
				\textbf{\textit{SP2UBI} (Proposed)} & \textbf{TFHE + Group Signatures + ISAC} & \ding{51} & \ding{51} & \ding{51} & \ding{51} & \ding{51} \\
				\bottomrule
				\multicolumn{7}{l}{\scriptsize \textbf{Legends:} \ding{51} Fully Supported / Resistant; \quad $\sim$ Partially Supported / Limited by operational constraints; \quad \ding{55} Not Supported / Vulnerable.} \\
				\multicolumn{7}{l}{\scriptsize \textsuperscript{a} Ensures the execution environment cannot deduce intermediate logical states or boolean comparison results.} \\
				\multicolumn{7}{l}{\scriptsize \textsuperscript{b} Mitigates the "Oracle Problem" by validating physical ground-truth independently of the vehicle's native internal sensors.} \\
				\multicolumn{7}{l}{\scriptsize \textsuperscript{c} Executes fraud validation or anomaly detection without necessitating partial data disclosure or privacy-invasive spot-checks.}
			\end{tabular}
		}
	\end{table*}

	\subsection{Comparative Evaluation with State-of-the-Art Schemes}
	To accurately delineate the functional position of \textit{SP2UBI}, Table \ref{tab:comparative_analysis} provides a systematic comparison with several prominent UBI frameworks. This evaluation is conducted across five critical dimensions: end-to-end (E2E) ciphertext confidentiality, physical layer tamper resistance, spatiotemporal unlinkability, conditional anonymity, and confidential anomaly detection.
	
	As illustrated in the table, existing literature shows a dichotomy between physical tamper resistance (mitigating the Oracle Problem) and strict data privacy. Early frameworks like \textit{PriPAYD}~\cite{Troncoso2007PriPAYDPP} achieve only a partial level of physical layer tamper resistance, as they rely heavily on the theoretical assumption of "tamper-resistant black boxes," an assumption vulnerable to pre-digitization signal injection in practice. Subsequent schemes, such as \textit{VPriv}~\cite{Popa2009VPrivPP} and \textit{DUBI}~\cite{Qi2021ScalableDP}, attempt to address this by incorporating roadside spot-checks or trusted third-party auditors. Consequently, they also only achieve a partial level of physical layer tamper resistance. In addition, such auditing mechanisms inherently undermine privacy by temporarily exposing the real identity and location of the vehicle during the verification phase. Conversely, recent architectures, including the frameworks proposed by Huang et al.~\cite{Huang2023BlockchainAssistedPC} and \textit{BE-VIP}~\cite{Sahu2024BlockchainAM}, achieve strong trajectory unlinkability via ZKPs and federated structures. However, they mathematically assume the digitized sensor data to be truthful, rendering them vulnerable to hardware spoofing. The \textit{SP2UBI} framework bridges this gap by leveraging the physical properties of ISAC waves (e.g., Doppler shifts) to securely verify the vehicle's kinematics without relying on onboard sensors or privacy-invasive spot-checks.
	
	Achieving confidential anomaly detection without compromising the E2E confidentiality of encrypted data remains an ongoing challenge. While \textit{PRIDE}~\cite{Wan2018PRIDEAP} ensures execution integrity, its reliance on intermediate plaintext thresholding (e.g., Boolean values for speeding anomalies) inadvertently leaks the temporal distribution of driver violations to all consensus nodes. Similarly, \textit{BE-VIP} executes machine learning risk models locally on plaintext within the OBU. This design necessitates the long-term retention of unencrypted telematics data at the endpoint, expanding the attack surface against hardware tampering. Even ZKP-based models like \textit{DUBI} and the Huang et al. framework often require partial data disclosure (e.g., opening cryptographic commitments) to trusted third-party auditors when anomalies trigger a dispute resolution process. Because these existing frameworks ultimately depend on endpoint plaintext processing (e.g., \textit{BE-VIP}), privacy-invasive physical audits (e.g., \textit{VPriv}), or partial data disclosure for resolving anomalies (e.g., \textit{DUBI} and Huang et al.), they offer only a limited guarantee in confidential anomaly detection. By utilizing the TFHE framework, \textit{SP2UBI} ensures data encryption at the source and guarantees that all subsequent arithmetic and logical operations, including anomaly validations, are executed exclusively in the ciphertext domain. This reduces the endpoint attack surface and protects raw telematics data as well as intermediate execution states from all intermediate nodes.
	
	\section{PERFORMANCE ANALYSIS}
	To validate the performance and assess the computational efficiency of the proposed framework, the protocol was implemented in Python on a local machine equipped with an Intel Core i7-1065G7 CPU @ 1.30GHz (4 cores) running Ubuntu 22.04 LTS. 
	
	The homomorphic circuits were constructed utilizing the \textit{Concrete} framework developed by Zama\footnote{Zama, ``Concrete: TFHE Compiler that converts Python programs into FHE equivalent,'' open-source FHE framework, available online: \url{https://github.com/zama-ai/concrete}, 2022.}. This library leverages PBS within the TFHE scheme, enabling the execution of exact discrete operations directly over ciphertexts while managing noise proliferation effectively. To evaluate the system's capacity for risk assessment within the encrypted domain, the \textit{Concrete ML} library was utilized. This framework facilitates the conversion of standard machine learning models into equivalent TFHE circuits through built-in quantization techniques. Through this setup, it is demonstrated that the proposed architecture possesses the computational flexibility to integrate homomorphic machine learning evaluations, maintaining a practical overhead tailored for VANETs.
	
	To train and validate the risk factor computation model, the machine learning framework was implemented using the telematics driving dataset provided by So et al. \cite{risks9040058}. This dataset encapsulates records from 100,000 distinct automobile insurance policies, capturing traditional risk variables and granular telematics features. To align the predictive model with the behavioral parameters of the protocol, a targeted feature mapping was executed. Specifically, the data series corresponding to aggressive driving maneuvers (namely hard braking (\texttt{Brake}) and rapid acceleration (\texttt{Accel})) were extracted and normalized alongside the total mileage driven and the percentage of vehicle operation during high-risk hours. The risk assessment was formulated as a linear regression problem executed entirely within the FHE domain via \textit{Concrete ML}. This specific linear formulation was deliberately chosen to satisfy the algebraic constraints of the decoupled HomMAC execution integrity protocol established in Section~\ref{subsec:hommac_temporal}.
	
	\subsection{Computational Overhead}
	The computational complexity imposed by the cryptographic operations executed at the OBU is systematically evaluated. Recognizing that the OBU performs both lattice-based FHE encryption and elliptic-curve-based group signatures, the TFHE lattice parameters were maintained constant (yielding fixed encryption times), while the signature generation was evaluated across various elliptic curve configurations. The resulting execution timings are summarized in Table~\ref{table:comp_times}.
	
	\begin{table}[htbp]
		\centering
		\caption{Computational Overhead Evaluation at the OBU}
		\label{table:comp_times}
		\setlength{\tabcolsep}{3pt} 
		\renewcommand{\arraystretch}{1.2}
		
		\resizebox{\columnwidth}{!}{%
			\begin{tabular}{|>{\centering\arraybackslash}m{2.2cm}|
					>{\centering\arraybackslash}m{3.2cm}|
					>{\centering\arraybackslash}m{3.2cm}|
					>{\centering\arraybackslash}m{2.0cm}|
					>{\centering\arraybackslash}m{1.8cm}|}
				\hline
				\textbf{Elliptic Curve Configuration} &
				\textbf{$t_{\mathsf{Enc}}(\llbracket v_{\text{sen}}^{(d)} \rrbracket_{\mathit{pk}_{\text{SV}}})$ (ms)} &
				\textbf{$t_{\mathsf{Enc}}(\llbracket \mathcal{P}^{(d)} \rrbracket_{\mathit{pk}_{\text{Ins}}})$ (ms)} &
				\textbf{$t_{\text{GroupSign}}$ (ms)} &
				\textbf{$t_{\text{OBU Total}}$ (ms)} \\ \hline
				
				SS512  & \multirow{5}{*}{2.38} & \multirow{5}{*}{7.14} & 16.087  & 25.607  \\ \cline{1-1} \cline{4-5}
				SS1024 &                       &                       & 198.460 & 207.980 \\ \cline{1-1} \cline{4-5}
				MNT159 &                       &                       & 29.728  & 39.248  \\ \cline{1-1} \cline{4-5}
				MNT201 &                       &                       & 39.181  & 48.701  \\ \cline{1-1} \cline{4-5}
				MNT224 &                       &                       & 48.086  & 57.606  \\ \hline
			\end{tabular}%
		}
	\end{table}
	
\begin{figure}[htbp]
		\centering
		\includegraphics[width=1\columnwidth]{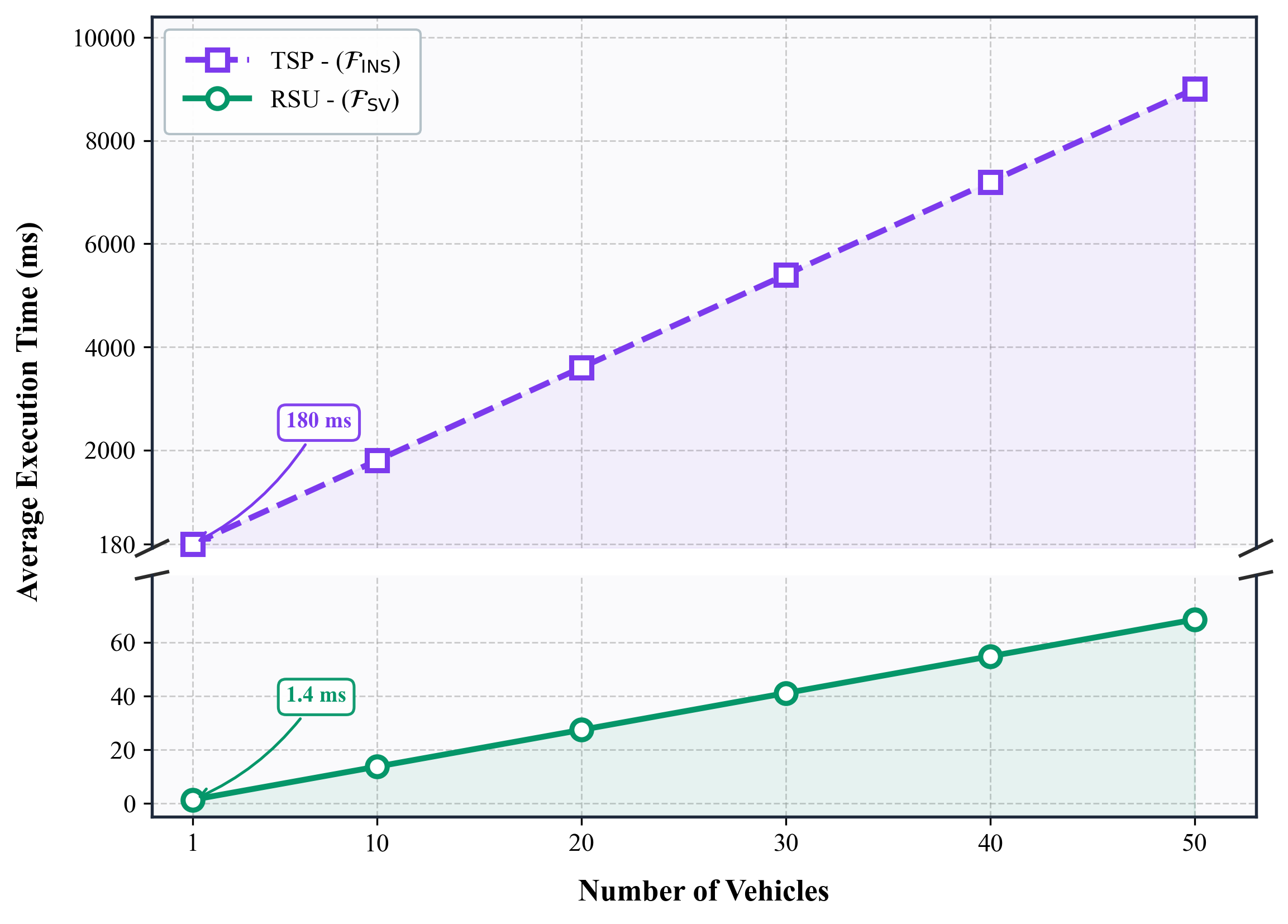}
		\caption{Computational overhead of infrastructure components (RSU and TSP).}
		\label{fig:infra_comp_times}
	\end{figure}

	The computational overhead incurred at the infrastructure layers is evaluated to ascertain the real-time feasibility of the protocol. As illustrated in Fig.~\ref{fig:infra_comp_times}, the execution time for the RSU involves a single, mixed-domain homomorphic subtraction circuit for speed verification, whereas the TSP's processing time for the baseline risk factor calculation accounts for 5 homomorphic multiplications and 4 homomorphic additions.
	
\begin{figure}[htbp]
		\centering
		\includegraphics[width=1\columnwidth]{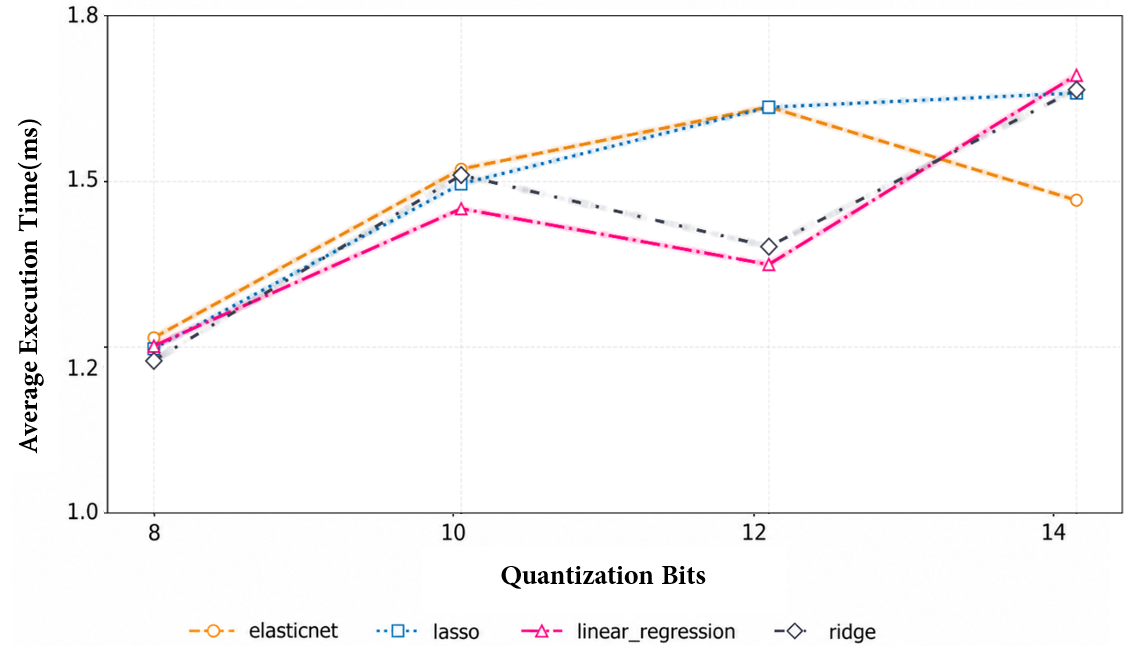}
		\caption{Computational overhead of FHE-based machine learning models for TSP risk factor calculation.}
		\label{fig:ml_comp_times}
	\end{figure}

	Fig.~\ref{fig:ml_comp_times} illustrates the computational overhead associated with integrating various linear regression algorithms (including standard linear regression, Ridge, Lasso, and ElasticNet) within the FHE framework. While in a standard FHE circuit implemented via \textit{Concrete}, model coefficients are explicitly encrypted as ciphertexts, learning-based risk assessment frameworks compiled through \textit{Concrete ML} apply optimized quantization and secure compilation to the underlying parameters. This cryptographic approach prevents the circuit evaluator (e.g., the TSP) from reconstructing the model's proprietary architecture or exact weights. Specifically, the critical quantization scale factors and zero-point parameters remain strictly private to the model owner, preserving the confidentiality of the insurer's model weights during domain-specific evaluations.
	
\begin{figure}[htbp]
		\centering
		\includegraphics[width=1\columnwidth]{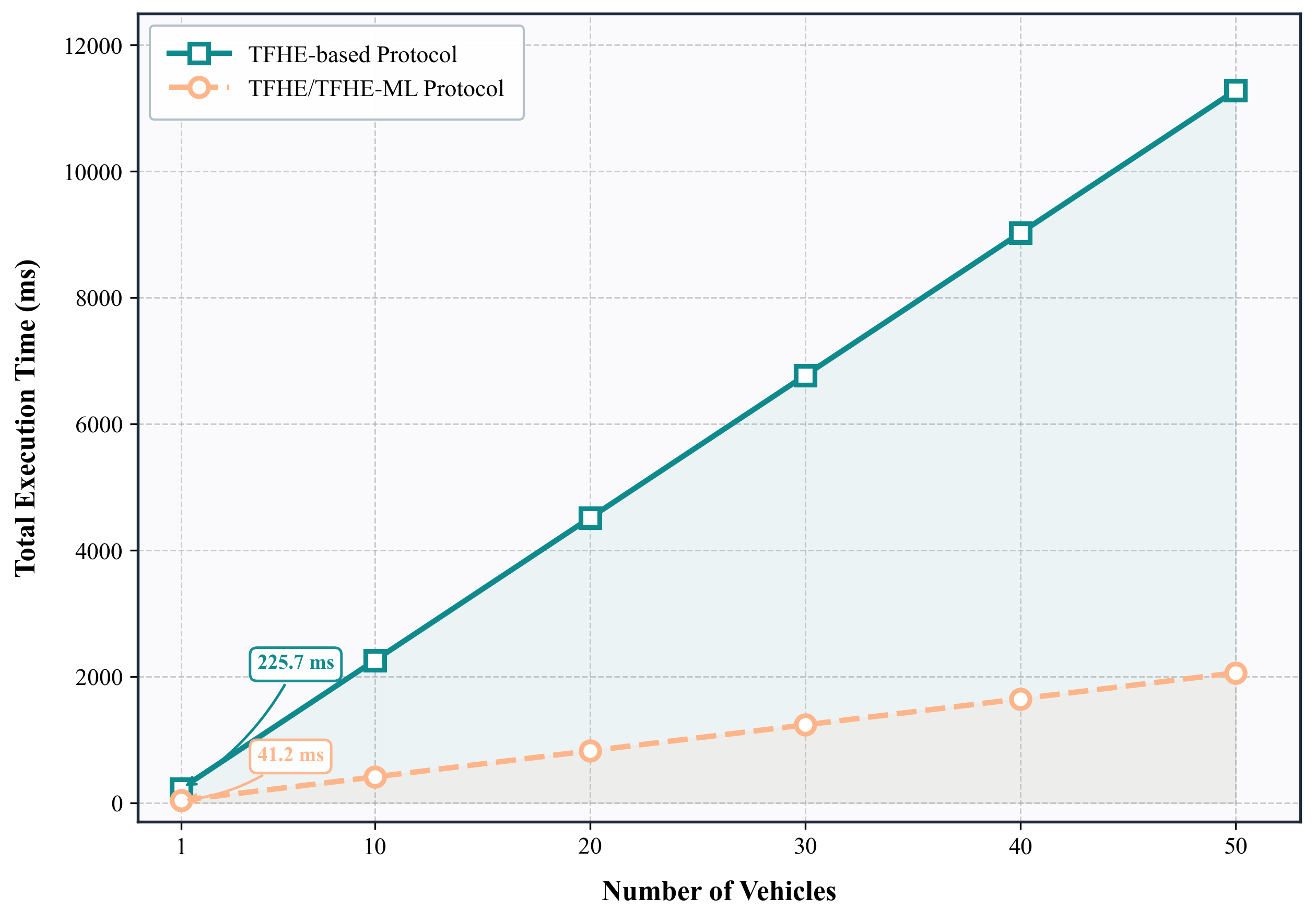}
		\caption{Total end-to-end execution time comparison.}
		\label{fig:total_execution_time}
	\end{figure}

	The end-to-end evaluation of the protocol is illustrated in Fig.~\ref{fig:total_execution_time}. The performance of the risk factor computation circuit is evaluated using both the conventional FHE-based approach and the TFHE-based linear machine learning models at the TSP.
	
	In the comparative evaluation, the proposed \textit{SP2UBI} scheme utilizes the TFHE framework, whereas the PRIDE scheme \cite{Wan2018PRIDEAP} relies on a matrix decomposition and multiplication encryption technique. PRIDE compromises end-to-end data privacy because the intermediate comparison outcomes (e.g., boolean values indicating speeding or acceleration thresholds) are directly revealed in plaintext within the smart contract execution environment. This fundamental cryptographic leakage exposes the chronological distribution of a driver's traffic violations to all consensus nodes. In contrast, \textit{SP2UBI} ensures data confidentiality throughout all processing phases without exposing intermediate states, while remaining computationally efficient and suitable for deployment in real-world scenarios.
	
	\subsection{Communication and Storage Overhead}
	Modern commercial OBUs, such as those manufactured by Cohda Wireless, typically feature at least $8\text{~GB}$ of internal storage. Consequently, the $4.888\text{~MB}$ storage overhead required by the proposed scheme over a 30-day epoch occupies $0.059\%$ of the available capacity, demonstrating its efficiency and practicality for continuous deployments.
	
	\begin{table}[htbp]
		\centering
		\caption{Storage Overhead at the OBU over a 30-Day Epoch}
		\label{table:vehicle_data_size}
		\setlength{\tabcolsep}{6pt}
		\renewcommand{\arraystretch}{1.1}
		
		\resizebox{\columnwidth}{!}{%
			\begin{tabular}{|>{\centering\arraybackslash}m{3.0cm}|
					>{\centering\arraybackslash}m{3.0cm}|
					>{\centering\arraybackslash}m{3.0cm}|}
				\hline
				\textbf{$\mathcal{S}(\llbracket \mathcal{P} \rrbracket)$ (MB)} &
				\textbf{$\mathcal{S}(\llbracket v_{\text{sen}} \rrbracket)$ (MB)} &
				\textbf{$\mathcal{S}_{\text{Total\_Veh}}$ (MB)} \\ \hline
				
				4.758 & 0.130 & 4.888 \\ \hline
			\end{tabular}%
		}
	\end{table}
	
\begin{figure}[htbp]
		\centering
		\includegraphics[width=1\columnwidth]{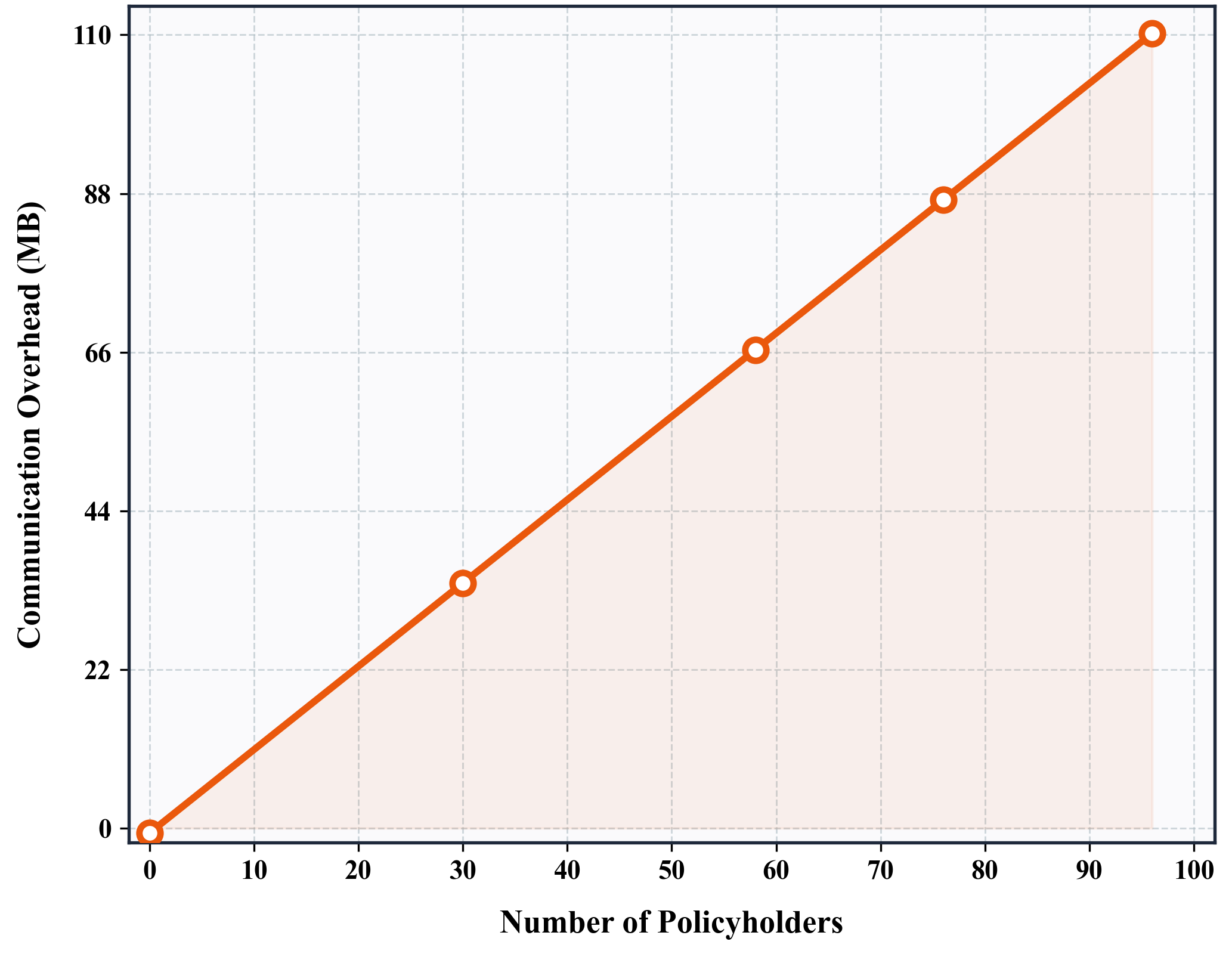}
		\caption{Communication overhead between the TSP and the Insurer.}
		\label{fig:tsp_ins_comm}
	\end{figure}

	The communication overhead for transmitting the daily payload between the OBU and the RSU is detailed in Table~\ref{table:obu_to_rsu_comm}, while the subsequent overhead for forwarding the evaluated batch between the RSU and the TSP is presented in Table~\ref{table:rsu_to_tsp_comm}. In addition, the communication overhead between the TSP and the insurer is illustrated in Fig.~\ref{fig:tsp_ins_comm}. These empirical results demonstrate that \textit{SP2UBI} maintains a low communication footprint, rendering it scalable and optimized for ITS.
	
	\begin{table}[htbp]
		\centering
		\caption{Communication Overhead from OBU to RSU}
		\label{table:obu_to_rsu_comm}
		\renewcommand{\arraystretch}{1.2}
		\resizebox{\columnwidth}{!}{%
			\begin{tabular}{|c|c|c|c|c|}
				\hline
				\textbf{$\text{Size}(\llbracket \mathcal{P}^{(d)} \rrbracket)$} & \textbf{$\text{Size}(\llbracket v_{\text{sen}}^{(d)} \rrbracket)$} & \textbf{$\text{Size}(\sigma^{(d)})$} & \textbf{$\text{Size}(\mathit{ID}_{\text{Ins}})$} & \textbf{$\text{Size}(\mathcal{C}^{(d)})$} \\
				\hline
				162.42 KB & 4.45 KB & 1.1 KB & 0.02 KB & 167.99 KB \\
				\hline
			\end{tabular}%
		}
		
		\vspace{1.5em}
		
		\caption{Communication Overhead from RSU to TSP}
		\label{table:rsu_to_tsp_comm}
		\renewcommand{\arraystretch}{1.2}
		\resizebox{0.65\columnwidth}{!}{%
			\begin{tabular}{|c|c|c|}
				\hline
				\textbf{$\text{Size}(\mathcal{R}_{\text{SV}}^{(d)})$} & \textbf{$\text{Size}(\mathcal{C}^{(d)})$} & \textbf{$\text{Size}(\mathit{RSU}_M^{(d)})$} \\
				\hline
				4.45 KB & 167.99 KB & 172.44 KB \\
				\hline
			\end{tabular}%
		}
	\end{table}
	
	\section{Conclusion}
	In this paper, \textit{SP2UBI}, a framework that performs verifiable risk assessments over homomorphically encrypted data, was proposed. By preventing the disclosure of raw driving metrics, the protocol mitigates the exposure of sensitive spatiotemporal trajectories. In addition, by compiling the insurer's risk evaluation algorithms into equivalent TFHE circuits, the framework maintains the confidentiality of the model weights and internal parameters from intermediate edge evaluators. The protocol also integrates a decoupled HomMAC architecture to verify mathematical execution integrity, while incorporating an ISAC-assisted speed validation mechanism to counter physical-layer sensor spoofing and resolve data-association ambiguities.
	
	Formal security and privacy evaluations, structured across sequential mathematical proofs, demonstrate that the architecture satisfies day-to-day unlinkability, $k$-anonymity against eavesdroppers, semantic security ($\mathit{IND\text{-}CPA}$), and source location obfuscation. Empirical performance evaluations show that these theoretical properties can be achieved with practical overheads tailored for vehicular networks. Specifically, the total on-board storage requirement over a 30-day billing epoch occupies less than $4.9\text{~MB}$, and the computational and communication footprints remain compatible with the real-time operational capacities of standard vehicular and roadside infrastructure. Future research will focus on extending this framework toward a privacy-preserving driver behavior management model for real-time safety monitoring.
	\bibliographystyle{IEEEtran}
	\bibliography{references}
	%
	
	
	%
	
	
	
	

\end{document}